\documentclass[letterpaper,twocolumn,prb,superscriptaddress]{revtex4-1}
\AtBeginDocument{
\heavyrulewidth=.08em
\lightrulewidth=.05em
\cmidrulewidth=.03em
\belowrulesep=.65ex
\belowbottomsep=0pt
\aboverulesep=.4ex
\abovetopsep=0pt
\cmidrulesep=\doublerulesep
\cmidrulekern=.5em
\defaultaddspace=.5em
}

\usepackage{amsmath,amssymb,graphicx}
\usepackage{physics}
\usepackage{float}
\usepackage[usenames,dvipsnames]{xcolor}
\usepackage{tikz,amsthm,comment}
\usepackage{booktabs}
\usepackage[export]{adjustbox}
\usepackage[section]{placeins}
\usepackage[percent]{overpic}
\usepackage{microtype}
\usepackage{array}
\usetikzlibrary{arrows.meta, automata, positioning, quotes, shapes}

\bibpunct{[}{]}{,}{n}{}{}
\usepackage[hidelinks]{hyperref} 
\hypersetup{colorlinks=true,citecolor=Red,linkcolor=Blue,urlcolor=Blue}

\newcommand{\xbond}{
	\begin{tikzpicture} [baseline={([yshift=-.5ex]current bounding box.center)}, scale=0.35]
		\begin{scope}[scale=0.3,
			every node/.style={anchor=west,
			draw,
			transform shape,
			minimum width=1cm,
			outer sep=0,
			},
			]
			\node[state, fill=black] (a) {};
			\node[state, fill=black] (b)[above right=1.0cm and 1.732cm of a] {};
			\node[state, densely dotted] (c1)[above=2.0cm of b] {};
			\node[state, densely dotted] (c2)[right=4cm of a] {};
			\node[state, densely dotted] (c3)[left=4cm of b] {};
			\node[state, densely dotted] (c4)[below=2cm of a] {};

			\draw[thick, color=black] (a) -- (b);
			\draw[dashed] (b) -- (c1);
			\draw[dashed] (b) -- (c2);
			\draw[dashed] (a) -- (c3);
			\draw[dashed] (a) -- (c4);
		\end{scope}
	\end{tikzpicture}
}
\newcommand{\zbond}{
	\begin{tikzpicture} [baseline={([yshift=-.5ex]current bounding box.center)}, scale=0.4]
		\begin{scope}[scale=0.3,
			every node/.style={anchor=west,
			draw,
			transform shape,
			minimum width=1cm,
			outer sep=0,
			},
			]
			\node[state, fill=black] (a) {};
			\node[state, fill=black] (b)[below=2cm of a] {};
			\node[state, densely dotted] (c1)[above right=1cm and 1.732cm of a] {};
			\node[state, densely dotted] (c2)[above left=1cm and 1.732cm of a] {};
			\node[state, densely dotted] (c3)[below right=1cm and 1.732cm of b] {};
			\node[state, densely dotted] (c4)[below left=1cm and 1.732cm of b] {};

			\draw[thick, color=black] (a) -- (b);
			\draw[dashed] (a) -- (c1);
			\draw[dashed] (a) -- (c2);
			\draw[dashed] (b) -- (c3);
			\draw[dashed] (b) -- (c4);
		\end{scope}
	\end{tikzpicture}
}

\newcommand{\honey}[4]{
	\def \w {white}
	\def \g {lightgray}
	\begin{tikzpicture}[baseline={([yshift=-.5ex]current bounding box.center)}, scale=0.5]
	  \begin{scope}[%
	every node/.style={anchor=west,
	regular polygon, 
	regular polygon sides=6,
	draw,
	minimum width=1cm,
	outer sep=0,
	},
	      transform shape]
	    \node (A)[fill=#1] 	    	        {};
	    \node (B)[fill=#2]  at (A.corner 5) {};
	    \node (C)[fill=#3]  at (A.corner 1) {};
	    \node (D)[fill=#4]  at (C.corner 5) {};
	  \end{scope}
	\end{tikzpicture}
}

\definecolor{ForestGreen}{HTML}{668000}
\definecolor{red1}{HTML}{FF4136}
\definecolor{green1}{HTML}{00802b}

\begin{document}


\title{Detection of long-range entanglement in gapped quantum spin liquids\\
by local measurements}

\author{Shi Feng}
\email[E-mail:]{feng.934@osu.edu}
\affiliation{Department of Physics, The Ohio State University, Columbus, Ohio 43210, USA}
\author{Yanjun He}
\email[E-mail:]{he.1578@osu.edu}
\affiliation{Department of Physics, The Ohio State University, Columbus, Ohio 43210, USA}
\author{Nandini Trivedi}
\email[E-mail:]{trivedi.15@osu.edu}
\affiliation{Department of Physics, The Ohio State University, Columbus, Ohio 43210, USA}

\date{\today}

\begin{abstract}
Topological order, reflected in long
range patterns of entanglement, is quantified by the
topological entanglement entropy (TEE) $\gamma$. We show that for gapped quantum spin liquids (QSL) it is possible to extract $\gamma$ using two-spin local correlators. 
We demonstrate our method for the gapped $\mathbb{Z}_2$ Kitaev spin liquid on a honeycomb lattice with anisotropic interactions. We show that the $\gamma = \log 2$ for $\mathbb{Z}_2$ topological order can be simply extracted from local two-spin correlators across two different bonds, with an accuracy comparable or higher than the Kitaev-Preskill construction. This implies that the different superselection sectors of $\mathbb{Z}_2$ gauge theory determined by global Wilson loop operators can be fully reflected locally in the matter majorana sector.    

\end{abstract}

\maketitle

\section{Introduction}
\begin{figure}[h]
	\centering
	\includegraphics[width=\columnwidth]{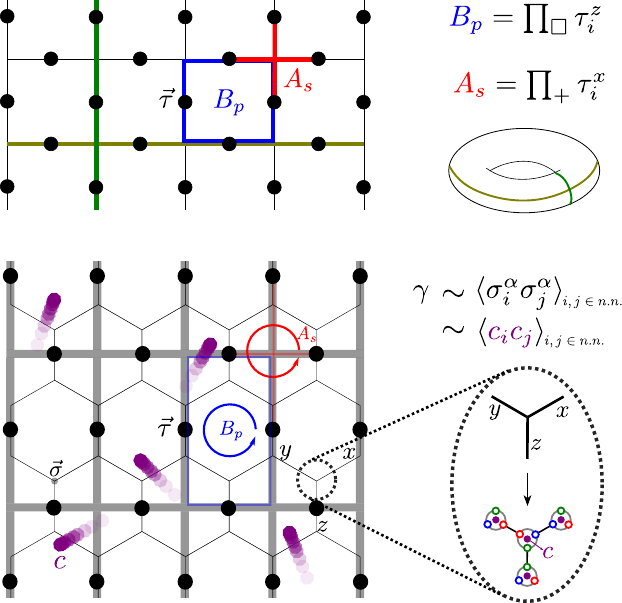}
	\caption{Top: TC model defined on the square lattice that gives $\mathbb{Z}_2$ gauge field. Two types of excitations $B_p$ and $A_s$ are marked in blue and red corresponding to that in Eq.\ref{eq:TC}; and two Wilson loops in corresponding to the big and small circles in the torus shown on the bottom right. $\gamma = \log 2$ is related to the eigenvalues of the two Wilson loops. Bottom: The effective TC model in the honeycomb lattice, where a spin of flavor $a$ is fractionalized into two majoranas $\sigma^a \sim ib^a c$. Here $c$ is the itinerant majorana sector (in purple) and $b^a$ is the localized majorana that constitutes the emergent $\mathbb{Z}_2$ gauge sector. The vortices on the plaquettes emergent from the gauge sector correspond to the TC excitation.     }
	\label{fig:TC}
\end{figure}

Quantum entanglement has started to play an increasingly important role in the understanding of quantum many-body systems. It is well-known for
its relation with topological order \cite{Wen1990,wen1995topological,Wen2002,Wen2010}, which is reflected in long range patterns of entanglement that are quantified by the topological entanglement entropy (TEE) \cite{Preskill2006}. One of the simplest examples of topological order is the emergent $\mathbb{Z}_2$ lattice gauge theory \cite{Wegner1971,Kogut1979,Senthil2000,Sandip2015} realized by the toric code (TC) model on a square lattice\cite{KITAEV20032}:
\begin{equation} \label{eq:TC}
    H = -\sum_{s} A_s - \sum_{p} B_p
\end{equation}
whose topological nature is reflected in the ground state degeneracy related to the different eigenvalues associated with Wilson loops around the torus shown in Fig.\ref{fig:TC} 

Usually, the extraction of TEE of a lattice gauge theory, like TC model, cannot be done by local operational protocol such as quantum distillation \cite{nielsen_chuang_2010,Sandip2015,Sandip2016}. The essential obstacle in calibrating entanglement by local operation is that the operators have to be locally gauge-invariant, which cannot detect other superselection sectors that therefore requires non-local operations. Also, from a statistical point of view, TEE is an intrinsic non-dyadic many-body correlation \cite{Matsuda2000,Preskill2006}, as is also reflected in the fact that any local correlation function in TC model with less than four qubits vanishes. Such non-dyadic nature directly implies that TEE cannot be extracted by one or two-point local measurements.  
Therefore the TEE in the $\mathbb{Z}_2$ lattice gauge field is usually extracted by a careful scaling analysis of von-Neumann entropy of the subsystem boundary length \cite{Furukawa2007,Jiang2012}; or by carefully engineered linear combinations of different subsystems such that non-topological contributions to the entanglement cancel \cite{Preskill2006, Wen2006}; or, for integrable models like TC, by exact derivation of $S_{vN}$ for large patches \cite{HAMMA200522,Hamma2005,Zeng2019,Castelnovo2014}. It is important to note that all of these methods are highly non-local operations on the many-body quantum system.

A different but related model, the Kitaev spin liquid defined on a honeycomb lattice with bond-dependent spin-spin interactions is exactly solvable. It is also described by a $\mathbb{Z}_2$ gauge theory but with matter majorana fermions \cite{KITAEV20062}. It shows a transition from a $\mathbb{Z}_2$ gapless quantum spin liquid (QSL) to a $\mathbb{Z}_2$ abelian gapped QSL as the strength of one of the bonds of the honeycomb lattice is increased compared to the other two \cite{KITAEV20062,Knolle2019,Blents1,ZhouRMP,Kasahara2018,Arakawa2021}. For large anisotropy of the bond strengths, the gapped Kitaev QSL (KSL) on a honeycomb lattice also maps to the TC on an underlying square lattice.

In this paper, we show as a proof of concept, that when a $\mathbb{Z}_2$ lattice gauge theory is embedded into the gapped QSL phase of Kitaev's honeycomb lattice (Fig.\ref{fig:TC}), the information of long-range entanglement can be fully encoded in local nearest-neighbor two-point correlators and it is therefore possible to extract TEE of the pure gauge theory from only a local measurement. Our central result for TEE is derived in terms of two-point correlators along a z-bond and an x-bond (or y-bond), as shown in Eq.\ref{tee-local} and schematically in Fig.\ref{fig:TC}. It is important to note that the two-point spin-spin correlators are given exactly by the local correlators of the matter majorana fermions. Since the matter majorana and gauge sectors are coupled \cite{Wen2003}, information about the $\mathbb{Z}_2$ non-local Wilson loops is imprinted in the \emph{local correlators of majoranas}.
Our results have important conceptual implications for the detection of long-range entanglement in topologically ordered systems.

\section{Toric Code in Honeycomb Lattice}
\begin{figure}[t]
	\centering
	\includegraphics[width=\columnwidth]{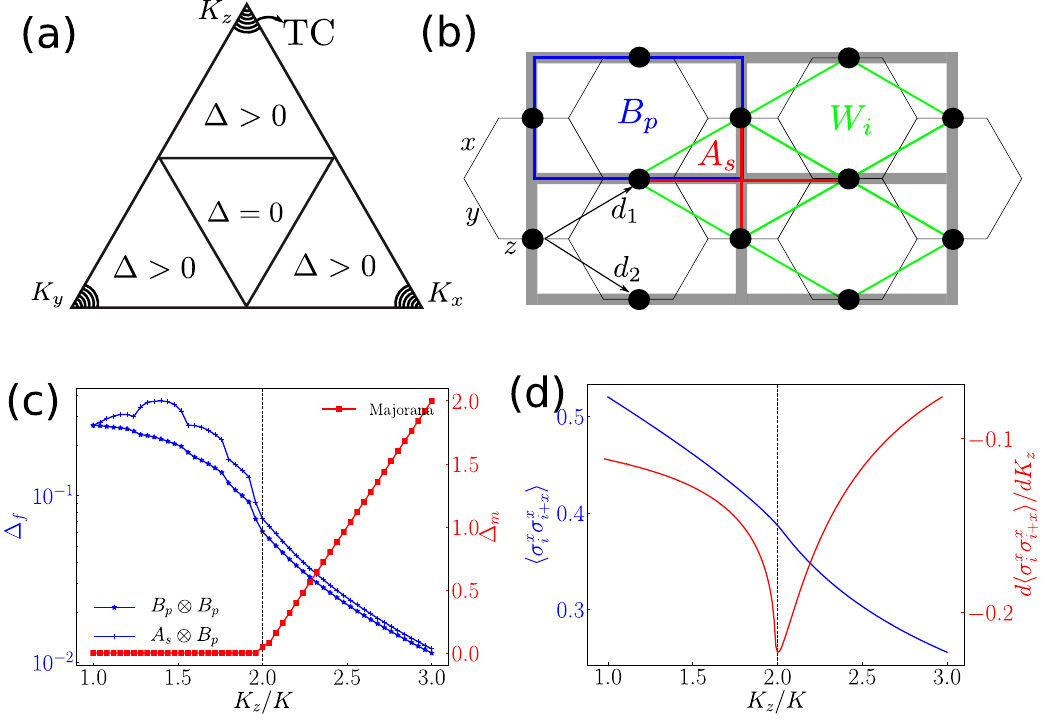}
	\caption{(a) The phase diagram of Kitaev model.  (b) The Kitaev honeycomb lattice. For large anisotropy $K_z/K > 2$,  the low energy physics is effectively captured by a square lattice where new spin-$\frac{1}{2}$ degrees of freedom $\tau$ (black bullets) live on the bonds of the honeycomb lattice, defining the Toric Code model. The excitations of the TC $A_s$ (e-charge), $B_p$ (m-charge) or the flux $W_i$ are marked by the colored plaquette and star in Fig.\ref{fig:TC}. (b) The evolution of energy gaps of different excitations in the Kitaev model as a function of anisotropy, shown on different scales. The phase transition from a gapless to a gapped QSL occurs at the $K_z/K=2$. Data are obtained by diagonalizing the majorana Hamiltonian at zero-flux sector with $28 \times 28$ unit cells. (c) Analytical results of the two-point correlation function and its derivative as a function of bond anisotropy, also shown on two different scales.}
	\label{fig:fig1}
\end{figure}

We begin with the mapping of the Kitaev honeycomb model to the TC model on a square lattice. 
The Hamiltonian of the Kitaev model is given by:
\begin{equation}
    H = K_x \sum_{\expval{ij}_x} \sigma^x_i \sigma^x_j  + K_y \sum_{\expval{ij}_y} \sigma^y_i \sigma^y_j + K_z \sum_{\expval{ij}_z} \sigma^z_i \sigma^z_j
\end{equation}
where $i,j$ label the sites of a hexagonal lattice, $\expval{ij}_a$ with $a=x,y,z$ denoting the nearest neighbor bonds in the $a-$th direction. It is known to have a gapless QSL at low anisotropy and a gapped $\mathbb{Z}_2$ QSL at $K_z/K > 2$ with $K = K_x = K_y$, as is shown in Fig.\ref{fig:fig1}(a,c,d). The gapped phase harbors abelian anyons and is connected to toric code gauge theory for $K_z/K \gg 2$ ~\cite{Wen2002,Adhip2021}. The toric code is defined on the effective square lattice as shown in Fig.\ref{fig:fig1}
\begin{equation}
H_{\text{TC}} = -J_{\text{TC}} \sum_p W_i,~~ W_i = \tau^z_{i+d_1}\tau^z_{i-d_2}\tau^y_i\tau^y_{i+d_1-d_2}
\end{equation}
with $J_{\text{TC}}=\frac{K^4}{16|K_z|^3}$ and $\tau^z = (\sigma_a^z-\sigma^z_b)/2$, where $a,b$ are sublattice indices and $d_1, d_2$ are lattice vectors. Upon performing a unitary rotation we have the familiar form of the TC Hamiltonian, given by
$    H_{\text{TC}} = -J_{\text{TC}}\left[\sum_{s} A_s+ \sum_{p} B_p\right]$, discussed above but now obtained as the large $K_z$ limit of the Kitaev honeycomb model.

For the gapped $\mathbb{Z}_2$ phase one can map $A_s$ and $B_p$ operators of the square lattice TC model to fluxes in the honeycomb model according to Fig.\ref{fig:fig1}(b) \cite{KITAEV20032}. 
Our aim here is to study the entanglement properties of $\mathbb{Z}_2$ topological order from the perspective of local measurements on the gapped $\mathbb{Z}_2$ KSL. As we demonstrate below, the correlation function between local degrees of freedom in the KSL is rich enough that non-trivial information is contained in local one and two-point density matrices; but also simple enough that only a few of all possible combinations of two-point correlators are needed to extract the TEE. Thus while TEE for the TC which is a pure gauge theory can only be obtained through non-local Wilson loops, the TEE for the KSL which has coupled matter and gauge degrees of freedom can be obtained by local correlators. 


\section{Local density matrix}
We start by constructing the reduced density matrix (RDM) of small subsystems. The smallest subsystem is a single spin. In the QSL ground state it has a  density matrix of size $2\times 2$ from which we obtain the local magnetization according to:
\begin{align} \label{eq:1qbit}
	\rho = \text{diag}(a, 1-a) + b\sigma^x + c\sigma^y,~
	\expval{\sigma^\alpha} &= \Tr(\rho \sigma^\alpha)
\end{align}
Since a QSL ground state necessarily has zero on-site magnetization along all axes, this immediately gives 
\begin{align}
a = &\frac{1+\expval{\sigma_z}}{2} = \frac{1}{2},~
c = \frac{\expval{\sigma_y}}{2}=0, ~
b = \frac{\expval{\sigma_x}}{2}=0  \\
&\Rightarrow~ \rho = \text{diag}\left(\frac{1}{2}, \frac{1}{2}\right) ~\Rightarrow~ S_{vN} = \log 2  
\end{align}
In order to construct a two-point RDM we need to determine all free parameters in the $4\times 4$ density matrix by measuring $\Tr(\rho \sigma^\alpha)$ and $\Tr(\rho \sigma_i^\alpha \sigma_j^\beta)$. 
In general, a two-point system defined on sites $i$ and $j$ can be captured by a 4$\times$4 RDM with $\Tr(\rho) = 1$ and $\rho^{\dag}=\rho$:
\begin{equation} \label{eq:rdm2}
    \rho = \frac{1}{4} \sum_{\alpha,\beta} \langle \sigma_i^\alpha  \sigma_j^\beta \rangle \sigma_i^\alpha \sigma_j^\beta,~\text{with }\alpha,\beta\in\{0,1,2,3\}
\end{equation}
where $\sigma^\alpha =(\mathbb{I}_2, \sigma^x,\sigma^y,\sigma^z)$ are the Pauli matrices and identity matrix and $\langle \sigma_i^\alpha  \sigma_j^\beta \rangle$ is the ground state expectation value of the corresponding operator $ \sigma_i^\alpha  \sigma_j^\beta$ for the two-point system. 

In the TC square lattice of $H_{\text{TC}}$ the two-point correlator is always zero \cite{Hamma2005} which gives a trivial RDM similar to that of the one-point system. However, it is non-zero when embedded in the honeycomb lattice whereby information of entanglement can be extracted.
The computation can be greatly simplified in the Kitaev model by noting that only the majorana sector contributes to local correlation functions \cite{Baskaran2007}. 
The correlation functions in the generalized Kitaev honeycomb model are highly anisotropic and extremely short ranged; the only non-zero correlators are $\expval*{\sigma_j^\alpha \sigma_{j+\beta}^\alpha}\delta_{\alpha,\beta}$, where $\sigma_{j+\beta}$ denotes the nearest-neighbor qubit connected by the $\beta$ bond. 
This can be made explicit if we separate an eigenstate into gauge and matter sectors, i.e. $\ket{\psi} = \ket{M_\mathcal{G},\mathcal{G}}$ with $\mathcal{G}$ denoting the $\mathbb{Z}_2$ gauge configuration and  $M_\mathcal{G}$ the matter majorana fermions on the gauge background. In this representation, the spin is fractionalized into majoranas $\sigma_j^\alpha = ib_j^\alpha c_j$, and the Hamiltonian in a particular $\ket{\mathcal{G}}$ sector becomes quadratic $H = i \sum_{\expval{ij}_\alpha} K_\alpha\: u_{\expval{ij}_\alpha} c_i c_j$, where $u_{\expval{ij}_a} = \pm 1$ gives the $\mathbb{Z}_2$ gauge field $\ket{\mathcal{G}}$ which constrains the configuration of gauge fluxes. This allows us to define the bond fermions $\eta_{\expval{ij}_\alpha} = \frac{1}{2}(b_i^\alpha + ib_j^\alpha)$ such that
\begin{equation} \label{eq:bondf}
	\sigma_i^\alpha = ic_i(\eta_{\expval{ij}_\alpha} + \eta_{\expval{ij}_\alpha}^\dagger),\;\;\; \sigma_j^\alpha = c_j (\eta_{\expval{ij}_\alpha}- \eta_{\expval{ij}_\alpha}^\dagger)
\end{equation}
where $i$ and $j$ belong to different sublattice. Note that the flux-free ground state corresponds to a full-filling of bond fermions; local spin-spin correlators can then be written as $\sigma_j^\alpha \propto c_j \hat{\pi}_{1,\expval{ij}_\alpha}\hat{\pi}_{2,\expval{ij}_\alpha}$ by Eq.(\ref{eq:bondf}), where $\hat{\pi}_{1,\expval{ij}_\alpha}$ and $\hat{\pi}_{2,\expval{ij}_\alpha}$ flip a pair of adjacent fluxes that share the same link $\expval{ij}_\alpha$. Then it can be readily seen that the only non-zero correlators are those sharing a link; those that don't share a link must vanish due to orthogonality of different flux configurations. Furthermore, only three nearest neighbor correlators ($\sigma_j^\alpha \sigma_{j+\alpha}^\alpha$ for $\alpha \in \{x,y,z\}$) give non-zero values because for $\sigma_j^\alpha \sigma_{j+\beta}^\alpha,~ \beta \neq \alpha$, we have
\begin{align} 
	\expval{\sigma_j^z \sigma_{j+z}^x} = \bra{\honey{\w}{\w}{\w}{\w};M_{\mathcal{G}}}ic_{j}c_{j+z}  \ket{M_{\mathcal{G}};\honey{\w}{\w}{\g}{\g}}
\end{align}
which must be zero due to orthogonality of states with and without $\mathbb{Z}_2$ flux. 
The remaining non-zero correlators can be calculated analytically in the $c$-majorana sector by
\begin{equation} \label{eq:corr}
    \langle \sigma_j^z \sigma_{j+z}^z \rangle = \expval{ic_j c_{j+z}} = \frac{\sqrt{3}}{16\pi^2} \int_{\text{BZ}} \frac{\epsilon_k}{\sqrt{\epsilon_k^2+\Delta_k^2}} d^2\vec{k}
\end{equation}
$\epsilon_k$ and $\Delta_k$ on the R.H.S. are defined by
\begin{align}
    \epsilon_k &= 2(K_x \cos k_1 + K_y \cos k_2+K_z)\\
    \Delta_k &= 2(K_x \sin k_1 + K_y \sin k_2)
\end{align}
where $k_1 = \bold{k} \cdot \bold{d}_1, ~k_2 = \bold{k} \cdot \bold{d}_2,$ and $\bold{d}_{1,2} = \frac{\sqrt{3}}{2} \bold{e}_y \pm \frac{1}{2} \bold{e}_x$ are unit vectors along $x$ and $y$ type bonds. The detailed derivation of Eq.\ref{eq:corr} as well as for the case with a weak TR-breaking perturbation are presented in the Appendix\ref{sec:der}. At the point $K_x = K_y = K_z$ we have $\langle \sigma_i^z \sigma_{i+z}^z \rangle = 0.52$; and as $K_z \rightarrow \infty$, we expect $\langle \sigma_i^z \sigma_{i+z}^z \rangle \rightarrow 1$. $ \langle \sigma_i^x \sigma_j^x \rangle$ and $ \langle \sigma_i^y \sigma_j^y \rangle$ can be obtained from Eq.\ref{eq:corr} by the substitutions $K_x \rightarrow K_z \rightarrow K_y \rightarrow K_x $ and $K_x \rightarrow K_y \rightarrow K_z \rightarrow K_x$, respectively.
From this information and the fact that QSL states have $\expval{\sigma_i^\alpha} = 0$, we can readily construct the two-point RDM for each type of dimer along the different bond directions. Let $\langle \sigma^{x}_i  \sigma^{x}_{i+x}  \rangle =4A$, $\langle \sigma^{y}_i  \sigma^{y}_{i+y}  \rangle =-4B$ and $\langle \sigma^{z}_i  \sigma^{z}_{i+z}  \rangle =1-4C$, by Eq.\ref{eq:rdm2} we have:
\begin{align}
    \rho_x &= \frac{1}{4}\mathbb{I}_4 + \frac{A}{4}\mathbb{J}_4\\
    \rho_y &= \frac{1}{4}\mathbb{I}_4 + \text{anti-diag}(B,-B,-B,B)\\
    \rho_z &= \text{diag}\left(\frac{1}{2}-C, C, C, \frac{1}{2}-C\right)
\end{align}
where $\mathbb{J}_4$ is the $4\times 4$ anti-diagonal unit matrix. These also show that the measurement of a local $\rho_\alpha$ only requires $\expval{\sigma^\alpha \sigma^\alpha}$ on an $\alpha$-type bond. The RDM of an $\alpha$-bond subsystem has two unique eigenvalues and two pairs of doubly degenerate eigenvectors, which immediately gives the von-Neumann entropy of an $\alpha$ bond: 
\begin{equation} \label{eq:svn}
	S_{vN}(\alpha) = -2\sum_\pm\lambda_\pm^\alpha\log(\lambda_\pm^\alpha),~ \lambda_\pm^\alpha = \frac{1\pm\expval*{\sigma^\alpha_{i}\sigma^\alpha_{i+\alpha}}}{4}
\end{equation}
Such degeneracy is directly a consequence of TR symmetry.
The two-point $\rho_\alpha$ has two pairs of two-fold degenerate eigenvalues if TR symmetry is present, and thus flux is conserved; any perturbation respecting the symmetry will not lift the degeneracy unless a phase transition occurs. 
Indeed, any perturbation to the Kitaev model which preserves TR symmetry leaves the system in the same spin liquid phase \cite{KITAEV20062}. As shown in Fig.\ref{fig:fig1}(d), the two-point correlators, and therefore the two-point RDM, directly reflects the majorana sector, and is able to detect the opening of the majorana gap at $K_z/K = 2$. Indeed, this directly explains the singular behavior of the local fidelity susceptibility at the phase transition reported in Ref.\cite{Wang2010}. But, as we show in the section below, there is more information in majorana sector that allows us to extract TEE from local correlators of the majorana particles.  


\section{Bipartite and topological entanglement entropy}
\begin{figure}[t]
	\centering
	\includegraphics[width=\columnwidth]{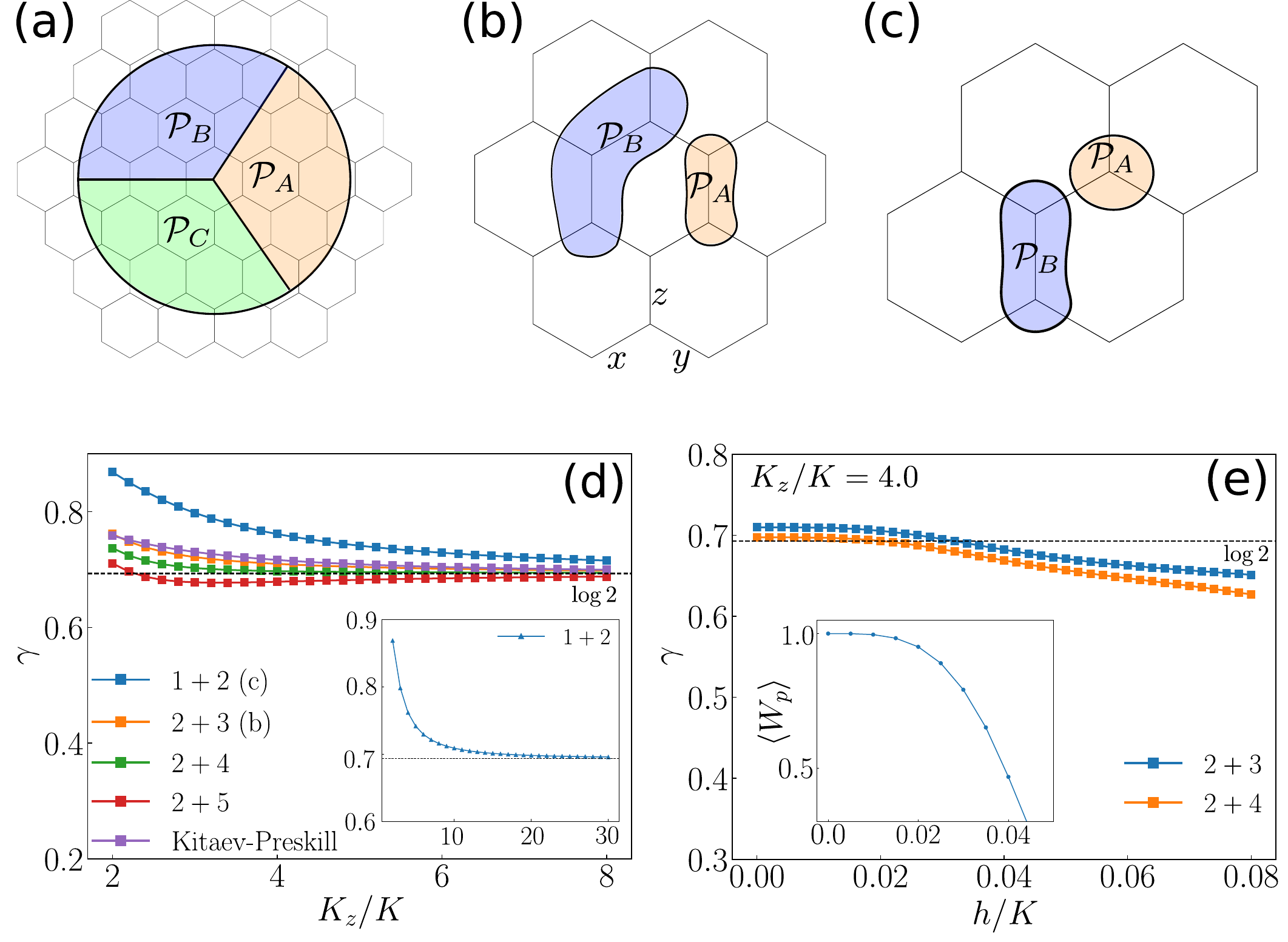}
	\caption{(a) Kitaev-Presill construction. (b) Partition scheme where subsystem $\mathcal{P}_A$ has two qubits living on a z bond, and $\mathcal{P}_B$ has three qubits living on a z bond and an x bond, and . (c) $\mathcal{P}_A$ as a single qubit subsystem, and $\mathcal{P}_B$ as two-point dimer on a z bond. (d) $\gamma$ extracted from different partition schemes using Eq.\ref{tee-local}, in comparison with that by the Kitaev-Preskill method. n+m in legends stands for n-site $\mathcal{P}_A$ and m-site $\mathcal{P}_B$, partly shown in the top row. The inset is a zoom-in near $\log2$ with a larger scope of $K_z$. (e) $\gamma$ extracted by the same method at $K_z/K = 4.0$ when the system is subjected to an out-of-plane magnetic field $h$. The inset shows the expectation of $W_p$ as an indicator of flux conservation.}
    	\label{fig:cuttings}
\end{figure}

In this section we show how to extract TEE using local measurements as defined in the previous section. This explicitly demonstrates that local degrees of freedom can contain information of non-local long-range entanglement. Thereafter we calculate arbitrary bipartite entanglement entropy from local measurements. 
In a gapped system the total von-Neumann entropy is written as
\begin{equation} \label{eq:Svn_all}
    S_{vN}(\mathcal{S}) = \alpha |\partial\mathcal{S}| - \gamma + O(e^{-|\partial\mathcal{S}|/\xi})
\end{equation}
where $\partial\mathcal{S}$ is the boundary of area $\mathcal{S}$, and $\xi$ the correlation length of the system. The first term in Eq.\ref{eq:Svn_all} is the non-topological area-law entropy.
Usually the TEE $\gamma$ is extracted by the Kitaev-Preskill construction shown in Fig.\ref{fig:cuttings}(a) as a linear combination of entropies of different subsystems: 
\begin{equation}
\begin{split}
        -\gamma = &S_{\mathcal{P}_A} + S_{\mathcal{P}_B} + S_{\mathcal{P}_C} \\ 
        & - S_{\mathcal{P}_{AB}} - S_{\mathcal{P}_{AC}} - S_{\mathcal{P}_{BC}} + S_{\mathcal{P}_{ABC}}
\end{split}
\end{equation}
with $\mathcal{S} \equiv \mathcal{P}_{ABC} = \mathcal{P}_A \cup \mathcal{P}_B \cup \mathcal{P}_C$. 
Indeed this is equivalent to a tripartite mutual information which is obtained by sampling multiple random variables of different subsystems. Hence it calibrates higher-order covariance and requires measurements on seven different subsystems.
The non-topological, non-area-law  $O(e^{-|\partial\mathcal{S}|/\xi})$ can depend on the specific partition and subsystem size $|\partial\mathcal{S}|$, but vanishes as $|\partial\mathcal{S}|\rightarrow \infty$. For computations on a finite lattice, subtracting area-law contribution from $S_{vN}$ may result in the calculated TEE $\tilde{\gamma}$ to deviate from the true $\gamma$.

However, we show that even though $\gamma$ is defined in Kitaev-Preskill construction as the information shared between three subsystems, it can nevertheless be extracted in a finite, or even two-point subsystem. This is possible if excitations relevant for short range entanglement are gapped enough, and the length scale of the interaction between (approximately) conserved gauge charges can be ignored. The gap $\Delta$ is inverse proportional to characteristic length scale $\xi$; for the gapped KSL under weak perturbation we effectively have $\xi \sim 1/\Delta \simeq 0$ for majoranas at high anisotropy, as is shown Fig.\ref{fig:fig1}(c). Also, the $\mathbb{Z}_2$ charges in the gapped KSL are conserved and do not interact with each other, hence there is no length scale associated with $\mathbb{Z}_2$ charges.  
Therefore the area-law and topological effects dominate over the $O(e^{-|\partial\mathcal{S}|/\xi})$, and the approximation $\tilde{\gamma} \simeq \gamma$ and $S_{vN} \simeq \alpha |\partial \mathcal{S}| - \gamma$ become accurate even in a finite subsystem with weak perturbation; we justify its applicability below with direct computation. 

We can formally decompose the non-topological entanglement of a multi-spin subsystem into several entangled pairs. In particular, we assume that the system is gapped enough and the short range entanglement remains significant up to next nearest neighbor spins. Then, the majority of the short range entanglement can be captured within a two-site subsystem. 
For a subsystem made of a z or x-bond dimer, we propose that the non-topological entanglement can be represented by bond-dependent entangled pairs separated across different bonds on the boundary shown pictorially:
\begin{equation} \label{eq:ansatz}
    \includegraphics[width=0.85\columnwidth]{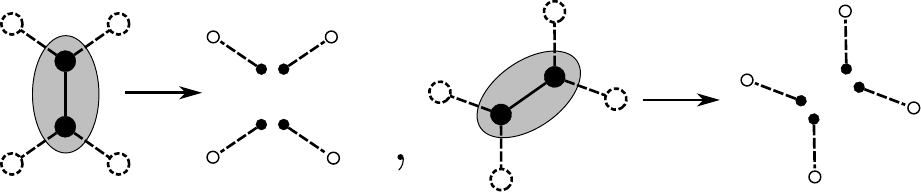} 
\end{equation}
Let $S_{vN}^\alpha$ denote the von-Neumann entropy of an entangled pair separated by an $\alpha$ bond, then the von-Neumann entropy is of the form
\begin{align}     
    S_{vN}\left(\zbond\right) &= 2S_{vN}^y + 2S_{vN}^x - \gamma \label{svn1}\\
    S_{vN}\left(\xbond\right) &= 2S_{vN}^y + 2S_{vN}^z - \gamma \label{svn2}
\end{align}
where the dotted circles denote relevant environmental degrees of freedoms. The structure of Eq.\ref{svn1} and Eq.\ref{svn2} is based on the lattice symmetry which describes the local contributions from the bonds that are cut on the boundary and the additional non-local topological contribution. The local contributions can be viewed as a “microscopic area law"; 
in addition, we include the topological contribution through $\gamma$ which cannot be described locally. Indeed, topological entropy is not subjected to quantum distillation or dilution into Bell pairs \cite{Sandip2015}. These equations immediately give
\begin{align} \label{eq:svn_x}
    S_{vN}^x &= \frac{1}{4}\left[S_{vN}\left(\zbond\right) + \gamma\right] = S_{vN}^y\\
    S_{vN}^z &= \frac{1}{2}S_{vN}\left(\xbond\right) - \frac{1}{4}S_{vN}\left(\zbond\right) + \frac{\gamma}{4} \label{eq:svn_z}
\end{align}
where we have exploited the $C_2^z$ symmetry. 
We would like to point out that in this representation, $S_{vN}^\alpha$ of entangled pairs is considered to have approximately incorporated the correction from $O(e^{-|\partial\mathcal{S}|/\xi})$ on a two-site scale, which is made explicit by the the formulation that entanglement entropy in R.H.S. are to be determined by the measured value of nearest neighbor correlators according to Eq.\ref{eq:svn}, instead of the \textit{a priori} area-law entropy. 

In particular in the TC limit $\abs{K_z /K} \rightarrow \infty$, we have $\expval{\sigma_i^z \sigma_{i+z}^z} \rightarrow 1$ and $\expval{\sigma_i^x \sigma_{i+x}^x} \rightarrow 0$, thus  $S_{vN}\left(\xbond\right) \rightarrow 2\log 2, ~S_{vN}\left(\zbond\right) \rightarrow \log 2$. This gives the von-Neumann entropy of entangled pairs of x and z type in a TC ground state: $S_{vN}^x(\text{TC}) = S_{vN}^y(\text{TC}) = \frac{1}{4}(\log 2 + \gamma)$, $S_{vN}^z(\text{TC}) = \frac{3}{4}\log 2 + \frac{1}{4}\gamma$,
which have both pair-wise contribution in the form of the scaled $\log2$,  and a topological contribution contained in $\frac{\gamma}{4}$. 

This picture provides a simple way to calculate entanglement entropy for arbitrary bipartite lattice, and simplifies the extraction of the topological entropy for applicable systems. Assuming a bipartite cutting 
 $\mathcal{S}\cup \mathcal{E}$ with $n_x$ x-bonds, $n_y$ y-bonds and $n_z$ z- bonds, the total von-Neumman entropy can be written as
\begin{equation} \label{eq:SvnS}
    S_{vN}(\mathcal{S}) = \sum_\alpha n_\alpha(\mathcal{S}) S_{vN}^\alpha(\gamma) - \gamma\  \cdot
\end{equation}
Its differential with respect to the subsystem $\mathcal{S}$ gives:
\begin{equation} \label{eq:deltasvn}
     \Delta S_{vN} = \sum_\alpha \Delta n_\alpha S_{vN}^\alpha(\gamma)
\end{equation}
where $\Delta S_{vN} = S_{vN}(\mathcal{S} + \Delta \mathcal{S}) - S_{vN}(\mathcal{S})$. The L.H.S. of Eq.\ref{eq:deltasvn} can be evaluated numerically in bipartite systems that are amenable to diagonalization, or calculated directly from single and two-point correlators for subsystems of less than two sites; and $\Delta n_\alpha$ is known from distinct subsystems in different partition schemes. One can readily extract the topological entropy $\gamma$ knowing the form of the local entropy contribution $S_{vN}^\alpha(\gamma)$ from the entangled pairs, which, in the case of Kitaev QSL, is given in Eq.\ref{eq:svn_x} and Eq.\ref{eq:svn_z}. These give the key equation for TEE in terms of two distinct two-point correlators:
\begin{equation}
    \begin{split}
        \gamma = &\frac{1}{\sum_\alpha \Delta n_\alpha} \Big[4\Delta S_{vN} - 2\Delta n_z S_{vN}\left(\xbond\right) \\
        &- (\Delta n_x + \Delta n_y - \Delta n_z) S_{vN}\left(\zbond\right)\Big]
    \end{split}
    \label{tee-local}
\end{equation}
Here one has the freedom to choose the area of each subsystem with different sets of $n_\alpha$. $S_{vN}$ can be computed as a contribution from bonds of using two-point correlators that invariably incorporate the correction from $O(e^{-|\partial\mathcal{S}|/\xi})$ and $\gamma$. 
Note that, in a QSL state, single-qubit entanglement entropy is unaffected by the short-range entanglement and $O(e^{-|\partial\mathcal{S}|/\xi})$ since it is fixed to be $\log 2$ by zero on-site magnetization. With this information, therefore, determining the entanglement entropy of another subsystem with more than one site suffices to extract $\gamma$ according to Eq.\ref{tee-local}. 

As an example, we now present this construction in the gapped $\mathbb{Z}_2$ QSL of Kitaev model using the scheme defined in Fig.\ref{fig:cuttings}(c), which requires only local measurements on a single qubit and a pair of two-point correlators on different bonds. Entanglement entropy for the z-bond dimer can be easily retrieved directly from the nearest-neighbor correlator from Eq.\ref{eq:svn}. 
In the TC limit, measurement of two-spin correlation gives $S_{vN}(\mathcal{P}_B) = S_{vN}\left(\zbond\right) = \log 2$ according to the previous discussion, and results in $\Delta S_{vN}(\mathcal{P}_A, \mathcal{P}_B) = 0$ where $\mathcal{P}_A$ is a single qubit; and by the same token $S_{vN}\left(\xbond\right) = 2\log 2$. After counting the number of boundary bonds for each cutting, Eq.\ref{tee-local} gives 
\begin{equation}
    \gamma(\text{TC}) = \log 2    
\end{equation}
which agrees exactly with the result derived from the Kitaev-Preskill construction for $L\rightarrow \infty$ and as well as with previously reported methods using large non-local partitions \cite{HAMMA200522,Hamma2005,Zeng2019} or non-trivial projections of wavefunction \cite{Castelnovo2014}. This exact derivation of $\gamma$ in TC limit relies on two facts (i) the majorana particles are highly gapped out, as is shown in Fig.\ref{fig:fig1}(c), whose relevant length scale vanishes accordingly; (ii) The conserved charges $A_s$ and $B_p$ do not interact with each other (unlike the non-Abelian phase which is discussed in Appendix.\ref{sec:der}), hence there is no length scale associated with the fluxes.  The alternative derivation of $\gamma(\text{TC}) = \log 2$ presented above makes explicit the intuition that topological order can be encoded even in local degrees of freedom, whose feature can be extracted by local two-point correlators. 
\section{Discussion}
\begin{figure}[t]
	\centering
	\includegraphics[width=\columnwidth]{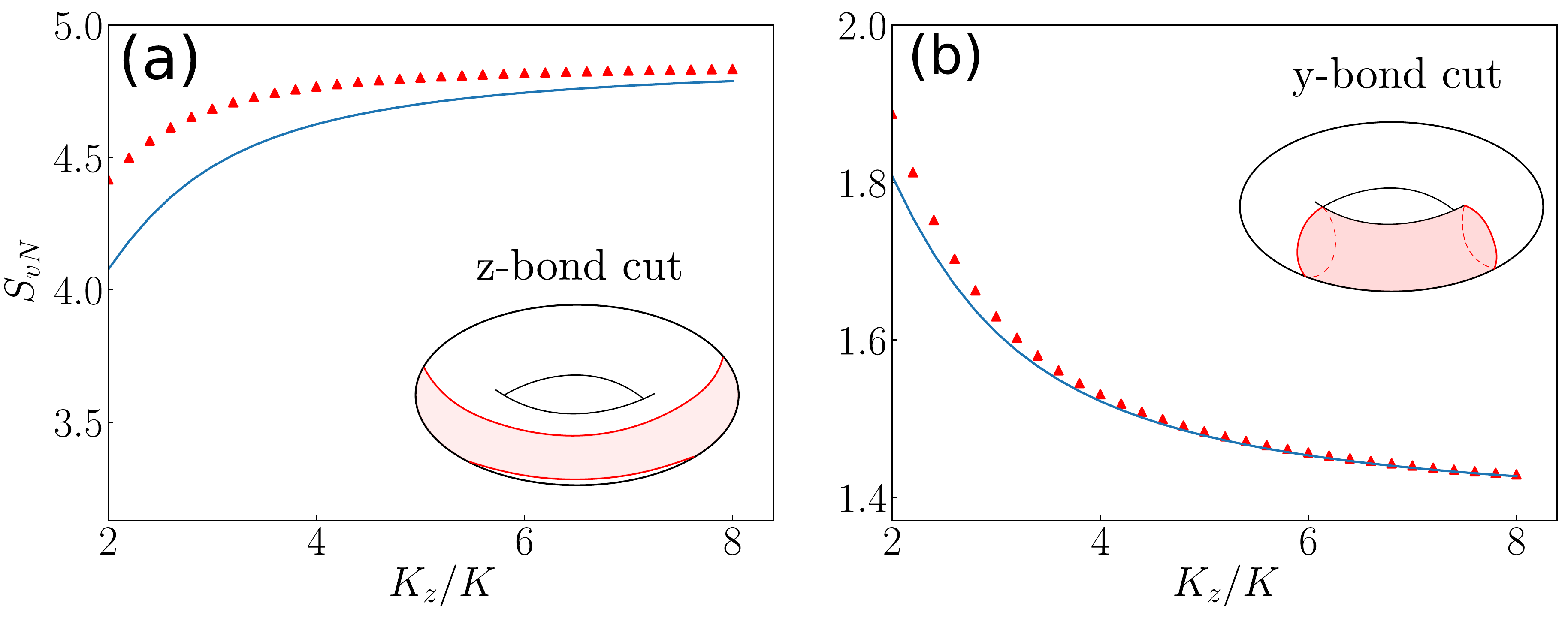}
	\caption{(a) Bipartite entanglement entropy of the subsystem enclosed by two z-bond cuts that traverse 8 z bonds. (b) Bipartite entanglement entropy of the subsystem enclosed by two y-bond cuts that traverse 6 y bonds. The solid blue lines show the results calculated by Eq.\ref{eq:svn_x}-Eq.\ref{eq:SvnS} with two-point correlator; and the red triangles are obtained numerically in 24-site lattice ($4\times3$ unit cells) with torus geometry.}
    	\label{fig:svnxy}
\end{figure}

To what extent is the construction in Eq.\ref{tee-local} applicable to QSL states with smaller majorana gaps and finite-size systems used in numerical diagonalization? To address this question, we numerically diagonalize a finite 24-site cluster of the Kitaev honeycomb Hamiltonian with different $K_z/K$ in the gapped $\mathbb{Z}_2$ phase. Figure.\ref{fig:cuttings}(d) shows $\gamma$ extracted by different methods, including the Kitaev-Preskill construction and our local-measurement method defined in Eq.\ref{tee-local} with some of the partitions shown in Fig.\ref{fig:cuttings}. These results provide a calibration of the applicability of the local-measurement method in comparison with the known analytical result. We demonstrate that it is possible to extract $\gamma$ by locally measuring single and two-point expectation values which is remarkably  capable of capturing the topological entropy. Furthermore, the accuracy improves as $K_z/K$ increases towards the TC limit where the gap $\Delta$ of majoranas becomes large. The exponentially suppressed error for large $K_z$ reflects the linear growth of the majorana gap as a function of $K_z$ as shown in Fig.\ref{fig:fig1}(c). Also, for a fixed $K_z/K$, the accuracy of extracted $\gamma$ can be greatly improved by slightly increasing the size of local subsystems, with comparable or smaller error than for $\gamma$ calculated by the Kitaev-Preskill construction which requires $S_{vN}$ of large patches. Knowing the topological entropy, it is straightforward to calculate arbitrary bipartite entanglement entropy using Eq.\ref{eq:svn_x}-Eq.\ref{eq:SvnS}, as shown in Fig.\ref{fig:svnxy}, where the calculated entropy agrees with that obtained by exact diagonalization for large anisotropy. 

It is worth pointing out that the construction is no longer accurate under larger TR-breaking perturbations whereby fluxes are not approximately conserved, or in the non-abelian phase with $0<K_z/K<2$ where fluxes interact with each other \cite{Lahtinen2012}. As is shown in Fig.\ref{fig:cuttings}(e), when the expectation of fluxes $W_p$ begins to deviate from unity, finite-range interaction between fluxes emerge, and non-topological contributions between gauge degrees of freedom to entanglement can contaminate the TEE.
\section{Conclusion}
In contrast to the TC model, for which the topological entanglement entropy is encoded in Wilson loops,  
making it unfeasible to extract TEE by local measurement, we show that the same is not true for the gapped Kitaev QSL on the honeycomb which allows for greater fine structure in terms of both matter and gauge sectors.
We find that by measuring local two-point correlators in the Kitaev model that target only the matter majorana fermions, it is possible to retrieve the exact topological entanglement entropy $\gamma = \log 2$ of the $\mathbb{Z}_2$ topological order. Indeed this implies that the emergent majorana particles contain the same topological information about emergent $\mathbb{Z}_2$ gauge field as is contained in the gauge sector.  
This construction is accurate away from the TC limit if $K_z/K$ is significantly larger than 2 and can be improved by increasing the subsystem size. 
The proposed construction remains valid under weak TR-breaking perturbation if fluxes are approximately conserved. 
This makes explicit the intuition that topological order, though being a non-local property, can be informed by local measurements.

Recently, randomized measurement protocols have been developed to measure entanglement and out-of-time-order correlators using two-point correlated noise spectroscopy,
including an experimental demonstration in a trapped-ion quantum simulator \cite{Lupke2020,Brydges2019,Vermersch2019}. There has also been a report of an experimental realization of the Toric code \cite{Satzinger2021,Verresen2021} using Rydberg atoms. We expect our ideas and calculations can give useful insights for extraction of TEE via local measurements in relevant experiments. 


\section*{Acknowledgements}
S.F. acknowledges support from NSF Materials Research Science and Engineering Center (MRSEC) Grant No. DMR-2011876, Y. H. from DOE grant DE-FG02-07ER46423, and N.T. from NSF-DMR 2138905. We thank Adhip Agarwala, Xu Yang and Yuanming Lu for discussions and comments.

\section*{Appendix}
\subsection{Two-point reduced density matrix} \label{sec:2qubit}
In general, a two-point system living on site $i$ and $j$ can be captured by a 4$\times$4 reduced density matrix (RDM) with $\Tr(\rho) = 1$ and $\rho^{\dag}=\rho$:
\begin{equation}
    \rho = \frac{1}{4} \sum_{i,j} \langle \sigma_i^\alpha  \sigma_j^\beta \rangle \sigma_i^\alpha \sigma_j^\beta,~\text{with },\alpha,\beta\in\{0,1,2,3\}
\end{equation}
where $\sigma^\beta =(\mathbb{I}_2, \sigma^x,\sigma^y,\sigma^z)$ are the Pauli matrices, and $\langle \sigma_i^\alpha  \sigma_j^\beta \rangle$ is the expectation value of the corresponding operator $ \sigma_i^\alpha  \sigma_j^\beta$ for the system.
Therefore such a density matrix consists of fifteen free real parameters (all the expectation values except $\langle \mathbb{I}_2  \mathbb{I}_2 \rangle $). 
If the two-points system is in a pure state, the second Renyi entropy will vanish:
\begin{equation}
    S_2 = -\log \Tr(\rho^2) = -\log\sum_{i,j} \langle \sigma_i^\alpha  \sigma_j^\beta \rangle^2  = 0
\end{equation}
This equation would eliminate one free parameter out of all these fifteen parameters, so we still have fourteen free parameters left. However, a pure state for a two-point system can be written as:
\begin{equation}
    \vert \Psi \rangle = a \vert 00 \rangle + b \vert 01 \rangle+c \vert 10 \rangle+d \vert 11 \rangle
\end{equation}
where $a,b,c,d$ are complex numbers satisfying the normalization condition. After eliminating the global phase factor by setting $a$ to be real and using normalization condition, there are only six free real parameters left for a pure two-points system. However, after applying the second Renyi entropy condition, there are fourteen free real parameters left in the reduced density matrix, and we still need eight equations to eliminate all the remaining free correlation functions to get only 6 free real parameters as in the wavefunction.

Indeed, there is a stronger condition than the second Renyi entropy to fully determine the pure state of a two-point system, that is the condition that the density matrix is a projector: $\rho^2 = \rho$. This constraint would further reduce the fifteen free parameters down to six, which can recover all the information of a pure state without redundancy. $\rho^2 = \rho$ gives 9 independent equations for all possible correlation functions of the two-point system (including $S_2 = 0$), so these 9 independent equations will provide the necessary direct checks on physical observables (all correlation functions) to see if a two-point system is pure.

However, $\rho^2 \neq \rho$ for reduced density matrices of mixed states, and those correlation functions don't have very clear relationships among themselves. Hence in principle it is always possible to determine if a system is entangled or not via local measurements. Furthermore, correlation functions contain more information about the underlying system than entanglement entropy alone, since the former also contains information of special geometry in its matrix elements which entanglement entropy, as a scalar, does not. This establishes the possibility that the correlation functions provide sharper resolution into the entanglement structure, so that one can distinguish the long-range topological entanglement from the non-topological ones by local measurements.  Therefore, as is shown in the main text  that a standalone correlator is relevant for TR-symmetry protected degeneracy and is able to detect phase transition between gapless and gapped QSL phases, but does not contain long-range entangled information; However, the combination of two local correlators can be used to extract topological entanglement entropy in the gapped $\mathbb{Z}_2$ QSL phase with very good accuracy.

\subsection{Derivation of correlation function in Kitaev model} \label{sec:der}
The Kitaev model can be split into gauge and matter sectors \cite{KITAEV20062}, i.e. $\ket{\psi} = \ket{M_\mathcal{G},\mathcal{G}}$ with $\mathcal{G}$ denoting the $\mathbb{Z}_2$ gauge configuration and  $M_\mathcal{G}$ the matter majorana fermions on the gauge background. In this representation, spin is fractionalized into majoranas $\sigma_j^a = ic_j b_j^a$, and the Hamiltonian in a particular $\ket{\mathcal{G}}$ sector becomes quadratic and integrable as  $H = i \sum_{\expval{ij}_a} K_a\: u_{\expval{ij}_a} c_i c_j$, where $u_{\expval{ij}_a} = \pm 1$ are good quantum numbers that determine $\ket{\mathcal{G}}$ by pinning down a particular configuration of gauge fluxes $\{W_p = \pm 1\}$. 

In an arbitrary eigenstate of the Kitaev Hamiltonian in some fixed gauge field configuration $\mathcal{G}$, we write the two-point spin correlation as
\begin{equation}
	\expval{\sigma_j^z(t) \sigma_{j+z}^z} = \bra{M_\mathcal{G}} \bra{\mathcal{G}} \sigma_i^a(t) \sigma_j^b(0) \ket{\mathcal{G}} \ket{M_\mathcal{G}}
\end{equation}
Let $i,j$ be on the same $z$ bond. Since fluxes in gauge sectors are conserved, with the majorana representation, the static two-spin correlation function for the bond $\expval{ij}_z$ becomes
\begin{equation} \label{eq:sjk}
	\begin{split}
		\expval{\sigma_j^z \sigma_{j+z}^z} = \bra{M_\mathcal{G}}i c_j c_k \ket{M_\mathcal{G}}
	\end{split}
\end{equation}
that is, the correlation is attributed to the matter fermion sector only. 
Recall that the zero-flux sector Hamiltonian is
\begin{equation}
\begin{split}
    	H &= \sum_{\mathbf{q}} 
	\begin{pmatrix}
		a_{-\mathbf{q}} & b_{-\mathbf{q}}
	\end{pmatrix}
	\begin{pmatrix}
		0 & if(\mathbf{q}) \\ -if^*(\mathbf{q}) & 0
	\end{pmatrix}
	\begin{pmatrix}
		a_\mathbf{q} \\ b_\mathbf{q}
	\end{pmatrix}
\end{split}
\end{equation}
or $H = \sum_{\mathbf{q}} \mathbf{\Psi}^\dagger \mathbf{h}(\mathbf{q}) \mathbf{\Psi}$, where $a$ and $b$ are momentum-space majorana operators on different sublattices. 
The off-diagonal elements for each majorana mode is related to $f(\mathbf{q}) = K_x e^{i\mathbf{q} \cdot \mathbf{n}_1} + K_y e^{i\mathbf{q}\cdot \mathbf{n}_2} + K_z \equiv  K_x e^{iq_x} + K_y e^{iq_y} + K_z $ where we've defined $q_x \equiv \mathbf{q}\cdot \mathbf{n}_1$ and $q_y \equiv \mathbf{q}\cdot \mathbf{n}_2$. Splitting its real and imaginary parts gives:
\begin{equation}
	f(\mathbf{q}) = \epsilon_\mathbf{q} + i\Delta_\mathbf{q}
\end{equation}
with
\begin{align}
	\epsilon_\mathbf{q} &= K_x \cos q_x + K_y\cos q_y + Jz\\
	\Delta_\mathbf{q} &= K_x\sin q_x + K_y \sin q_y
\end{align}
To diagonalize the Hamiltonian, note the block matrix can be written as $	\mathbf{h}(\mathbf{q}) = \mathbf{d}\cdot \vec{\sigma},\;\;\;\text{with } \mathbf{d} = (d_x, d_y, 0) = (-\Delta_\mathbf{q}, -\epsilon_q)$, 
so the eigen energy is just $E_{\mathbf{q},\pm} = \pm \sqrt{\abs{f}^2} = \pm \sqrt{\epsilon_\mathbf{q}^2 + \Delta_\mathbf{q}^2}$. 
Let $E_\mathbf{q} = E_{\mathbf{q},+} > 0$, the diagonalized Hamiltonian becomes
\begin{equation}
	H
	= \sum_{\mathbf{q}} E_\mathbf{q} \left( C_{\mathbf{q},1}^\dagger C_{\mathbf{q},1} - C_{\mathbf{q},2}^\dagger C_{\mathbf{q},2} \right) 
\end{equation}
where $C_{q,1},~ C_{q,2}$ are operators for the lower and upper complex majorana band respectively. 
Therefore the groud state is given by filling the lower band majorna:
\begin{equation} \label{eq:psi0}
	\ket{\Psi_0} = \prod_\mathbf{q} C_{\mathbf{q},2}^\dagger\ket{0} 
\end{equation}
It is straightforward to find:
\begin{equation}\label{eq:aqbq}
    \begin{split}
      	C_{\mathbf{q},2} &= \frac{1}{\sqrt{2}}\left(\frac{\sqrt{\epsilon_\mathbf{q}^2 + \Delta_\mathbf{q}^2}}{\Delta_\mathbf{q} - i\epsilon_\mathbf{q}}a_{\mathbf{q}} + b_{\mathbf{q}}\right),\\
	C_{-\mathbf{q},2} &= \frac{1}{\sqrt{2}}\left(-\frac{\sqrt{\epsilon_\mathbf{q}^2 + \Delta_\mathbf{q}^2}}{\Delta_\mathbf{q} + i\epsilon_\mathbf{q}}a_{-\mathbf{q}} + b_{-\mathbf{q}} \right)
    \end{split}
\end{equation}
where we used the fact that $\Delta_\mathbf{q}$ being anti-symmetric while $\epsilon_\mathbf{q}$ is symmetric. 
From Eq.\ref{eq:aqbq} the two-point majorana correlator in momentum space can be written as
\begin{equation} \label{eq:cjck}
    \begin{split}
        ic_j c_k &= \frac{i}{N} \sum_{\mathbf{q}} a_{-\mathbf{q}} b_{\mathbf{q}} + a_{\mathbf{q}} b_{-\mathbf{q}} \\
        &= \frac{1}{N} \sum_{\mathbf{q}}  \frac{\epsilon_\mathbf{q} - i\Delta_\mathbf{q}}{2E_\mathbf{q}} \Big( C_{-\mathbf{q},2} C_{\mathbf{q},2} \\
        &\phantom{==}- C_{\mathbf{q},2}^\dagger C_{-\mathbf{q},2}^\dagger + C_{-\mathbf{q},2} C_{-\mathbf{q},2}^\dagger - C_{\mathbf{q},2}^\dagger C_{\mathbf{q},2} \Big) \\
         &\phantom{==}+\text{h.c.}
    \end{split}
\end{equation}
Then by Eq.\ref{eq:sjk}, Eq.\ref{eq:psi0} and Eq.\ref{eq:cjck} we have
\begin{equation}
	\expval{\sigma_j^z \sigma_{j+z}^z} = \frac{1}{N}\sum_{\mathbf{q}\in \text{BZ}} \frac{\epsilon_\mathbf{q}}{E_\mathbf{q}} 
\end{equation}
In order to transform the correlator to Fourier space, we define the unit vectors of the lattice $\mathbf{n}_1 = (\frac{\sqrt{3}}{2}, \frac{1}{2}),\;\;
\mathbf{n}_2 =  (\frac{\sqrt{3}}{2}, - \frac{1}{2})$ 
with $\abs{\mathbf{n}_i} = 1$ and the corresponding reciprocal lattice vectors $\mathbf{b}_1 = \frac{2\pi}{\sqrt{3}}(1, \sqrt{3}),\;\;
\mathbf{b}_2 = \frac{2\pi}{\sqrt{3}}(1, -\sqrt{3})$. 
Note that $N$ is the total number of sites (number of unit cells is $N/2$), we can rewrite the sum into integral by $\frac{1}{N} \sum_{\mathbf{q} \in \text{BZ}} = \frac{1}{N\delta q}\int_{\text{BZ}} d^2\mathbf{q}$
where $\delta q$ is the volumn per allowed $q$. $\delta q$ is related to the real space volume $V$ by $\delta q = \frac{4\pi^2}{V} = \frac{16\pi^2}{\sqrt{3}N}$,
where we used $	V = \#u \times \abs{\mathbf{n}_1 \times \mathbf{n}_2} = \frac{N}{2} \times \frac{\sqrt{3}}{2} = \frac{\sqrt{3}}{4}N$, 
with $\# u = N/2$ the number of unit cells.
So we finally have
\begin{equation}
	\expval{\sigma_j^z \sigma_{j+z}^z} = \frac{\sqrt{3}}{16\pi^2}\int_{\text{BZ}} \frac{\epsilon_\mathbf{q}}{E_\mathbf{q}}d^2\mathbf{q}
\end{equation}
as shown in the main text (See also Ref.\cite{Baskaran2007}). 

\begin{figure}
    \centering
    \includegraphics[width=\columnwidth]{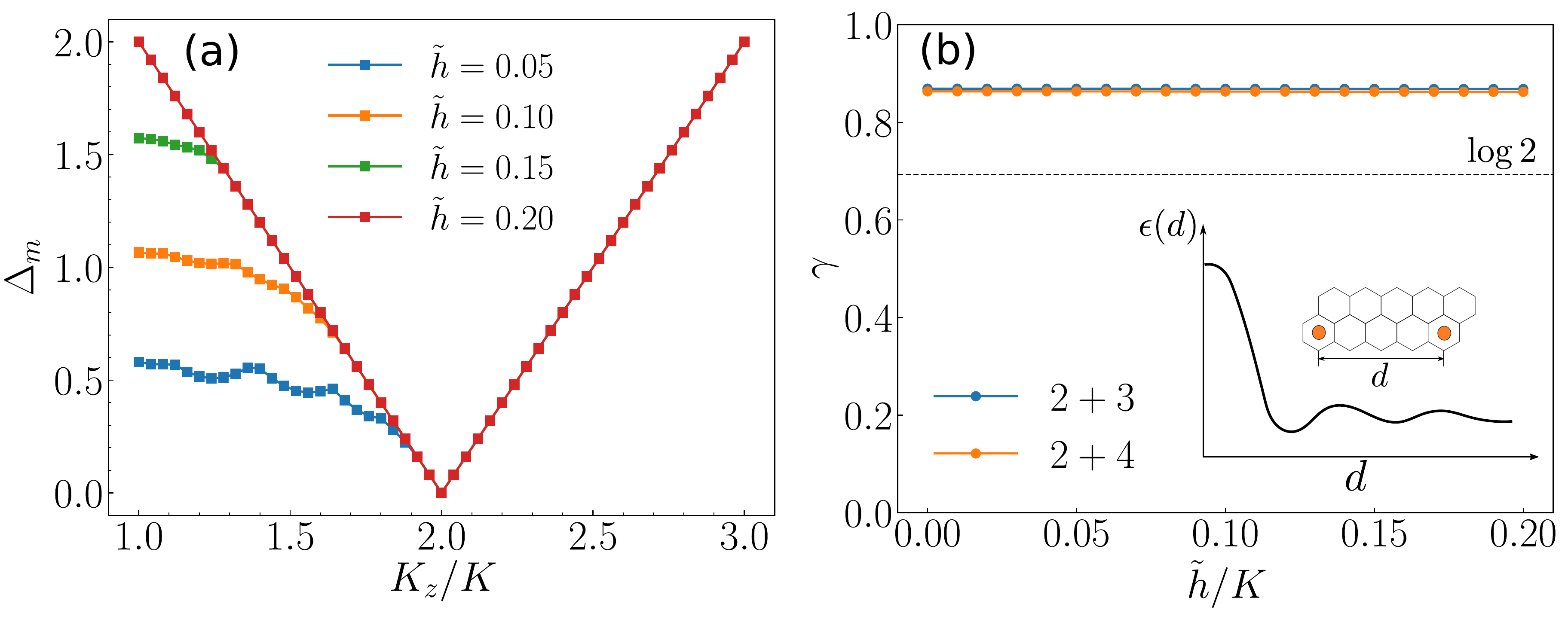}
    \caption{(a) The gap of itinerant majoranas at different anisotropy and TR-breaking perturbation $\tilde{h}$ that preserves flux conservation. In the non-Abelian phase the gap increase with $\tilde{h}$ but is upper-bounded by $\Delta_m = 2$. (b) $\gamma$ extracted by 2+3 and 2+4 construction introduced in the main text. The accuracy does not improve with increasing $\tilde{h}$ since the fluxes do interact with each other which introduces a length scale and short range entanglement. The inset is a schematic plot of energy splitting of two fluxes separated by a distance $d$. Detailed calculation are reported in Ref.\cite{Lahtinen2012}.   }
    \label{fig:app1}
\end{figure}

In the case of broken TR-symmetry, to leading order, the Hamiltonian takes the form \cite{KITAEV20062}:
\begin{equation}
	H = \sum_{\mathbf{q}}
	\begin{pmatrix}
		a_{-\mathbf{q}} & b_{-\mathbf{q}}
	\end{pmatrix}
	\begin{pmatrix}
		D(\mathbf{q}) & if(\mathbf{q}) \\
		-i f^*(\mathbf{q}) & D(\mathbf{q})
	\end{pmatrix}
	\begin{pmatrix}
		a_{\mathbf{q}} \\ b_{\mathbf{q}}
	\end{pmatrix}
\end{equation}
whose eigenvalues are $\pm E(\mathbf{q}) = \pm \sqrt{\abs{f(\mathbf{a})}^2 + D(\mathbf{q})^2}$. 
This immediately gives:
\begin{align}
    C_{\mathbf{q},1} &= \frac{1}{\sqrt{2}} \left( a_\mathbf{q} + \frac{D - if - E}{D + if^* - E}b_{\mathbf{q}} \right)\\
	C_{\mathbf{q},2} &= \frac{1}{\sqrt{2}} \left( \frac{D + if^* + E}{D - if + E} a_{\mathbf{q}} + b_{\mathbf{q}} \right)
\end{align}
In the ground state, only the lower band is occupied, so we need only to focus on $C_{\mathbf{q},2}$ and its conjugate operator. For simplicity we define $P(\mathbf{q}) \equiv \frac{D + if^* + E}{D - if + E}$ 
for $C_{\mathbf{q},2}$, hence we have
\begin{align}
	a_{\mathbf{q}} &= \frac{\sqrt{2}}{P(\mathbf{q}) - P^*(-\mathbf{q})} \left( C_{\mathbf{q},2} - C_{-\mathbf{q},2}^\dagger \right),\\
	b_{\mathbf{q}} &= -\frac{\sqrt{2}P^*(\mathbf{q})}{P(\mathbf{q}) - P^*(-\mathbf{q})}C_{\mathbf{q},2} \\
	&+ \sqrt{2}\left( 1 + \frac{P^*(\mathbf{q})}{P(\mathbf{q}) - P^*(-\mathbf{q})} \right) C_{-\mathbf{q},2}^\dagger  \nonumber
\end{align}
then the two-majorana correlator becomes:
\begin{equation}
	\begin{split}
		\expval{ic_j c_k} &= \frac{i}{N} \sum_{\mathbf{q}} \expval{a_{-\mathbf{q}} b_{\mathbf{q}} + a_\mathbf{q}b_{-\mathbf{q}}} \\
		 &\simeq \frac{\sqrt{3}}{16\pi^2}\int_{\text{BZ}} \frac{2 \left[ P^*(\mathbf{q}) + P^*(-\mathbf{q}) \right]  d^2\mathbf{q} }{\left[ P(-\mathbf{q}) - P^*(\mathbf{q}) \right] \left[ P(\mathbf{q}) - P^*(-\mathbf{q}) \right]}
	\end{split}
\end{equation}
This result, however, cannot be used to retrieve the TEE in the non-Abelian phase of Kitaev spin liquid. The reason is two-fold. Even though the majoranas are gapped out while retaining the conservation of fluxes, the gap of these itinerant majoranas are bounded hence there is an finite lower bound of $\xi$ which contradicts the assumption that $\xi \rightarrow 0$, as is shown in Fig.\ref{fig:app1}(a).  
Moverover, even though the fluxes in the non-Abelian phase are conserved, they still interact with each other so that the energy band is not flat, as shown in Fig.\ref{fig:app1}(b), which introduces an interaction length scale \cite{Lahtinen2012} and contaminates the TEE by finite range entanglement, adding an additional contribution to the local two-point measure. 

\subsection{Degeneracy of Entanglement Spectrum} \label{sec:deg}
In the Kitaev model, the degeneracy of the entanglement spectrum is protected by TR symmetry, or by the conservation of $\mathbb{Z}_2$ flux, or both.
To make this point explicit, 
consider a TR-symmetric two-point density matrix $\rho_2$, where $|\mathcal{S}| = 2$. It can be represented in its diagonal basis as
\begin{equation}
    \tilde{\rho}_2 = U \rho_2 U^\dagger = \text{diag}(a,b,c,d)
\end{equation}
where the tilde is used to denote the diagonal representation. Upon tracing out one of the qubits, either the first one or the second one, we obtain,
\begin{align}
        \Tr_{\mathcal{E} = 1'}(\tilde{\rho}_2) &= (a+b) \ket{\tilde{0}}\bra{\tilde{0}} + (c+d) \ket{\tilde{1}}\bra{\tilde{1}} = \tilde{\rho}_1 \\
        \Tr_{\mathcal{E} = 1}(\tilde{\rho}_2) &= (a+c) \ket{\tilde{0}}\bra{\tilde{0}} + (b+d) \ket{\tilde{1}}\bra{\tilde{1}} = \tilde{\rho}_{1'}
\end{align}
According to TR symmetry, the one-qubit density matrix must be $\tilde{\rho}_1 = \tilde{\rho}_{1'} = \text{diag}(\frac{1}{2}, \frac{1}{2})$, hence we must have $b=c$ and $a=d$. Therefore, $\rho_2$ will have two pairs of two-fold degenerate eigenvalues if TR symmetry is present; and any perturbation respecting the symmetry will not lift the degeneracy unless a phase transition occurs. Indeed, any perturbation to Kitaev model which preserves TR symmetry leaves the system in the same spin liquid phase. 


In fact, the degeneracy is present in a subsystem of arbitrary size. Consider a bipartite system $ \mathcal{S}\cup \mathcal{E}$, where  $\mathcal{S}$ includes $n$ particles. 
The reduced density matrix of the $n$-point subsystem is  
\begin{equation}
	\rho_n = \Tr_{i\in \mathcal{E}} \ket{\psi}\bra{\psi}
\end{equation}
The ground state $\ket{\psi}$ of Kitaev model respects time-reversal (TR) symmetry such that $\ket{\psi} = \Theta\ket{\psi}$,  where
\begin{equation}
    \Theta = \otimes_{i=1}^N \Theta_i,~~\Theta_i = \exp(-i\frac{\pi}{2}\sigma_i^y) \mathcal{K}
\end{equation}
with $\mathcal{K}$ denotes complex conjugation, $\Theta_i^2 = -1$ (fermionic condition) and $\Theta_i \Theta_i^\dagger = 1$. This is true regardless of fixing a particular gauge sector since $\Theta$ commutes with the Wilson loop operator. 
The $n$-qubit RDM of a Kitaev ground state can then be written as
\begin{equation}
	\begin{split}
		\rho_n &= \Tr_{i\in\mathcal{E}}\ket{\psi}\bra{\psi} =  \Tr_{i\in\mathcal{E}}\Theta\ket{\psi}\bra{\psi}\Theta^\dagger \\
		       &= \Theta_{\mathcal{S}}^{\phantom{\dagger}} \Tr_{i\in \mathcal{E}}\left(\Theta_{\mathcal{E}}^{\phantom{\dagger}} \ket{\psi}\bra{\psi}\Theta_{\mathcal{E}}^\dagger\right) \Theta_{\mathcal{S}}^\dagger = \Theta_{\mathcal{S}}^{\phantom{\dagger}} \rho_n \Theta_{\mathcal{S}}^\dagger
	\end{split}
\end{equation}
where $\Theta_{\mathcal{S}} \equiv \otimes_{i\in \mathcal{S}} \Theta_i$ and $\Theta = \Theta_\mathcal{S}\otimes\Theta_\mathcal{E}$. Hence $[\Theta_{\mathcal{S}},~\rho_n] = 0$.
Therefore the eigenvalues of RDM can be identified by the eigenvalues of $\Theta_{\mathcal{S}}$. Assume $n\in\text{odd}$ and $\ket{v}$ is an arbitrary eigenvector such that $\rho_n \ket{v} = v\ket{v}$. Then we must also have $	\rho_n [\Theta_{\mathcal{S}} \ket{v}] = v[\Theta_\mathcal{S}\ket{v}], ~\Theta_{\mathcal{S}}\ket{v} \perp \ket{v}$ due to $\Theta_\mathcal{S}^2 = -1$, 
so $v$ must be a two-fold degenerate eigenvalue according to Kramers' theorem. Therefore, there is a double degeneracy of the entire entanglement spectrum of $\rho_{n\in \text{odd}}$ which characterizes its topological nature \cite{Pollmann2010}. Such degeneracy of $\rho_{n\in \text{odd}}$ remains intact in the presence of perturbations that respect TR symmetry.  
This also holds for $n=1$ for a single qubit subsystem: in order to respect the degeneracy, the reduced density matrix $\rho_1$ must represent a maximally entangled state, so that $\rho_1 = \text{diag}(\frac{1}{2}, \frac{1}{2})$ and $S_{vN}(\rho_1) = \log 2$. 
However, for $n\in \text{even}$ this argument generally does not hold, for Kramers' theorem does not apply to even number of qubits whereby $\Theta_\mathcal{S} = 1$. 
Yet as we see in the two-fold degeneracy of $\rho_{x,y,z}$, the degeneracy is still present due to TR symmetry in the absence of the fermionic condition. Indeed, it is straightforward to prove by induction that such degeneracy is present for both $n\in\text{even}$ and $n\in\text{odd}$.  

In the absence of time-reversal (TR) breaking perturbations, the short-range contamination of the long-range entanglement of the $\mathbb{Z}_2$ KSL is attributed solely to the majorana sector, which, interestingly, turns out to be a strength for extracting the TEE, and exhibits a two-fold degeneracy of the two-point RDM reflecting the emergent $\mathbb{Z}_2$ gauge field. 
In particular, extraction of TEE from local correlators is not viable in the TC lattice where the two-point correlation matrix vanishes \cite{KITAEV20032,Hamma2005}; in contrast, the non-zero correlation in the gapped $\mathbb{Z}_2$ phase of KSL, which is attributed only to the majorana sector, can through local measurements inform the long-range entanglement of $\mathbb{Z}_2$ topological order that is attributed to the gauge sector.

\bibliography{reference.bib}

\begin{thebibliography}{37}%
\makeatletter
\providecommand \@ifxundefined [1]{%
 \@ifx{#1\undefined}
}%
\providecommand \@ifnum [1]{%
 \ifnum #1\expandafter \@firstoftwo
 \else \expandafter \@secondoftwo
 \fi
}%
\providecommand \@ifx [1]{%
 \ifx #1\expandafter \@firstoftwo
 \else \expandafter \@secondoftwo
 \fi
}%
\providecommand \natexlab [1]{#1}%
\providecommand \enquote  [1]{``#1''}%
\providecommand \bibnamefont  [1]{#1}%
\providecommand \bibfnamefont [1]{#1}%
\providecommand \citenamefont [1]{#1}%
\providecommand \href@noop [0]{\@secondoftwo}%
\providecommand \href [0]{\begingroup \@sanitize@url \@href}%
\providecommand \@href[1]{\@@startlink{#1}\@@href}%
\providecommand \@@href[1]{\endgroup#1\@@endlink}%
\providecommand \@sanitize@url [0]{\catcode `\\12\catcode `\$12\catcode
  `\&12\catcode `\#12\catcode `\^12\catcode `\_12\catcode `\%12\relax}%
\providecommand \@@startlink[1]{}%
\providecommand \@@endlink[0]{}%
\providecommand \url  [0]{\begingroup\@sanitize@url \@url }%
\providecommand \@url [1]{\endgroup\@href {#1}{\urlprefix }}%
\providecommand \urlprefix  [0]{URL }%
\providecommand \Eprint [0]{\href }%
\providecommand \doibase [0]{http://dx.doi.org/}%
\providecommand \selectlanguage [0]{\@gobble}%
\providecommand \bibinfo  [0]{\@secondoftwo}%
\providecommand \bibfield  [0]{\@secondoftwo}%
\providecommand \translation [1]{[#1]}%
\providecommand \BibitemOpen [0]{}%
\providecommand \bibitemStop [0]{}%
\providecommand \bibitemNoStop [0]{.\EOS\space}%
\providecommand \EOS [0]{\spacefactor3000\relax}%
\providecommand \BibitemShut  [1]{\csname bibitem#1\endcsname}%
\let\auto@bib@innerbib\@empty
\bibitem [{\citenamefont {Wen}\ and\ \citenamefont {Niu}(1990)}]{Wen1990}%
  \BibitemOpen
  \bibfield  {author} {\bibinfo {author} {\bibfnamefont {X.~G.}\ \bibnamefont
  {Wen}}\ and\ \bibinfo {author} {\bibfnamefont {Q.}~\bibnamefont {Niu}},\
  }\href {\doibase 10.1103/PhysRevB.41.9377} {\bibfield  {journal} {\bibinfo
  {journal} {Phys. Rev. B}\ }\textbf {\bibinfo {volume} {41}},\ \bibinfo
  {pages} {9377} (\bibinfo {year} {1990})}\BibitemShut {NoStop}%
\bibitem [{\citenamefont {Wen}(1995)}]{wen1995topological}%
  \BibitemOpen
  \bibfield  {author} {\bibinfo {author} {\bibfnamefont {X.-G.}\ \bibnamefont
  {Wen}},\ }\href {\doibase 10.1080/00018739500101566} {\bibfield  {journal}
  {\bibinfo  {journal} {Advances in Physics}\ }\textbf {\bibinfo {volume}
  {44}},\ \bibinfo {pages} {405} (\bibinfo {year} {1995})}\BibitemShut
  {NoStop}%
\bibitem [{\citenamefont {Wen}(2002)}]{Wen2002}%
  \BibitemOpen
  \bibfield  {author} {\bibinfo {author} {\bibfnamefont {X.-G.}\ \bibnamefont
  {Wen}},\ }\href {\doibase 10.1103/PhysRevB.65.165113} {\bibfield  {journal}
  {\bibinfo  {journal} {Phys. Rev. B}\ }\textbf {\bibinfo {volume} {65}},\
  \bibinfo {pages} {165113} (\bibinfo {year} {2002})}\BibitemShut {NoStop}%
\bibitem [{\citenamefont {Chen}\ \emph {et~al.}(2010)\citenamefont {Chen},
  \citenamefont {Gu},\ and\ \citenamefont {Wen}}]{Wen2010}%
  \BibitemOpen
  \bibfield  {author} {\bibinfo {author} {\bibfnamefont {X.}~\bibnamefont
  {Chen}}, \bibinfo {author} {\bibfnamefont {Z.-C.}\ \bibnamefont {Gu}}, \ and\
  \bibinfo {author} {\bibfnamefont {X.-G.}\ \bibnamefont {Wen}},\ }\href
  {\doibase 10.1103/PhysRevB.82.155138} {\bibfield  {journal} {\bibinfo
  {journal} {Phys. Rev. B}\ }\textbf {\bibinfo {volume} {82}},\ \bibinfo
  {pages} {155138} (\bibinfo {year} {2010})}\BibitemShut {NoStop}%
\bibitem [{\citenamefont {Kitaev}\ and\ \citenamefont
  {Preskill}(2006)}]{Preskill2006}%
  \BibitemOpen
  \bibfield  {author} {\bibinfo {author} {\bibfnamefont {A.}~\bibnamefont
  {Kitaev}}\ and\ \bibinfo {author} {\bibfnamefont {J.}~\bibnamefont
  {Preskill}},\ }\href {\doibase 10.1103/PhysRevLett.96.110404} {\bibfield
  {journal} {\bibinfo  {journal} {Phys. Rev. Lett.}\ }\textbf {\bibinfo
  {volume} {96}},\ \bibinfo {pages} {110404} (\bibinfo {year}
  {2006})}\BibitemShut {NoStop}%
\bibitem [{\citenamefont {Wegner}(1971)}]{Wegner1971}%
  \BibitemOpen
  \bibfield  {author} {\bibinfo {author} {\bibfnamefont {F.~J.}\ \bibnamefont
  {Wegner}},\ }\href {\doibase 10.1063/1.1665530} {\bibfield  {journal}
  {\bibinfo  {journal} {Journal of Mathematical Physics}\ }\textbf {\bibinfo
  {volume} {12}},\ \bibinfo {pages} {2259} (\bibinfo {year}
  {1971})}\BibitemShut {NoStop}%
\bibitem [{\citenamefont {Kogut}(1979)}]{Kogut1979}%
  \BibitemOpen
  \bibfield  {author} {\bibinfo {author} {\bibfnamefont {J.~B.}\ \bibnamefont
  {Kogut}},\ }\href {\doibase 10.1103/RevModPhys.51.659} {\bibfield  {journal}
  {\bibinfo  {journal} {Rev. Mod. Phys.}\ }\textbf {\bibinfo {volume} {51}},\
  \bibinfo {pages} {659} (\bibinfo {year} {1979})}\BibitemShut {NoStop}%
\bibitem [{\citenamefont {Senthil}\ and\ \citenamefont
  {Fisher}(2000)}]{Senthil2000}%
  \BibitemOpen
  \bibfield  {author} {\bibinfo {author} {\bibfnamefont {T.}~\bibnamefont
  {Senthil}}\ and\ \bibinfo {author} {\bibfnamefont {M.~P.~A.}\ \bibnamefont
  {Fisher}},\ }\href {\doibase 10.1103/PhysRevB.62.7850} {\bibfield  {journal}
  {\bibinfo  {journal} {Phys. Rev. B}\ }\textbf {\bibinfo {volume} {62}},\
  \bibinfo {pages} {7850} (\bibinfo {year} {2000})}\BibitemShut {NoStop}%
\bibitem [{\citenamefont {Ghosh}\ \emph {et~al.}(2015)\citenamefont {Ghosh},
  \citenamefont {Soni},\ and\ \citenamefont {Trivedi}}]{Sandip2015}%
  \BibitemOpen
  \bibfield  {author} {\bibinfo {author} {\bibfnamefont {S.}~\bibnamefont
  {Ghosh}}, \bibinfo {author} {\bibfnamefont {R.~M.}\ \bibnamefont {Soni}}, \
  and\ \bibinfo {author} {\bibfnamefont {S.~P.}\ \bibnamefont {Trivedi}},\
  }\href {\doibase 10.1007/JHEP09(2015)069} {\bibfield  {journal} {\bibinfo
  {journal} {Journal of High Energy Physics}\ }\textbf {\bibinfo {volume}
  {2015}},\ \bibinfo {pages} {69} (\bibinfo {year} {2015})}\BibitemShut
  {NoStop}%
\bibitem [{\citenamefont {Kitaev}(2003)}]{KITAEV20032}%
  \BibitemOpen
  \bibfield  {author} {\bibinfo {author} {\bibfnamefont {A.}~\bibnamefont
  {Kitaev}},\ }\href {\doibase https://doi.org/10.1016/S0003-4916(02)00018-0}
  {\bibfield  {journal} {\bibinfo  {journal} {Annals of Physics}\ }\textbf
  {\bibinfo {volume} {303}},\ \bibinfo {pages} {2} (\bibinfo {year}
  {2003})}\BibitemShut {NoStop}%
\bibitem [{\citenamefont {Nielsen}\ and\ \citenamefont
  {Chuang}(2010)}]{nielsen_chuang_2010}%
  \BibitemOpen
  \bibfield  {author} {\bibinfo {author} {\bibfnamefont {M.~A.}\ \bibnamefont
  {Nielsen}}\ and\ \bibinfo {author} {\bibfnamefont {I.~L.}\ \bibnamefont
  {Chuang}},\ }\href {\doibase 10.1017/CBO9780511976667} {\emph {\bibinfo
  {title} {Quantum Computation and Quantum Information: 10th Anniversary
  Edition}}}\ (\bibinfo  {publisher} {Cambridge University Press},\ \bibinfo
  {year} {2010})\BibitemShut {NoStop}%
\bibitem [{\citenamefont {Soni}\ and\ \citenamefont
  {Trivedi}(2016)}]{Sandip2016}%
  \BibitemOpen
  \bibfield  {author} {\bibinfo {author} {\bibfnamefont {R.~M.}\ \bibnamefont
  {Soni}}\ and\ \bibinfo {author} {\bibfnamefont {S.~P.}\ \bibnamefont
  {Trivedi}},\ }\href {\doibase 10.1007/JHEP01(2016)136} {\bibfield  {journal}
  {\bibinfo  {journal} {Journal of High Energy Physics}\ }\textbf {\bibinfo
  {volume} {2016}},\ \bibinfo {pages} {136} (\bibinfo {year}
  {2016})}\BibitemShut {NoStop}%
\bibitem [{\citenamefont {Matsuda}(2000)}]{Matsuda2000}%
  \BibitemOpen
  \bibfield  {author} {\bibinfo {author} {\bibfnamefont {H.}~\bibnamefont
  {Matsuda}},\ }\href {\doibase 10.1103/PhysRevE.62.3096} {\bibfield  {journal}
  {\bibinfo  {journal} {Phys. Rev. E}\ }\textbf {\bibinfo {volume} {62}},\
  \bibinfo {pages} {3096} (\bibinfo {year} {2000})}\BibitemShut {NoStop}%
\bibitem [{\citenamefont {Furukawa}\ and\ \citenamefont
  {Misguich}(2007)}]{Furukawa2007}%
  \BibitemOpen
  \bibfield  {author} {\bibinfo {author} {\bibfnamefont {S.}~\bibnamefont
  {Furukawa}}\ and\ \bibinfo {author} {\bibfnamefont {G.}~\bibnamefont
  {Misguich}},\ }\href {\doibase 10.1103/PhysRevB.75.214407} {\bibfield
  {journal} {\bibinfo  {journal} {Phys. Rev. B}\ }\textbf {\bibinfo {volume}
  {75}},\ \bibinfo {pages} {214407} (\bibinfo {year} {2007})}\BibitemShut
  {NoStop}%
\bibitem [{\citenamefont {Jiang}\ \emph {et~al.}(2012)\citenamefont {Jiang},
  \citenamefont {Wang},\ and\ \citenamefont {Balents}}]{Jiang2012}%
  \BibitemOpen
  \bibfield  {author} {\bibinfo {author} {\bibfnamefont {H.-C.}\ \bibnamefont
  {Jiang}}, \bibinfo {author} {\bibfnamefont {Z.}~\bibnamefont {Wang}}, \ and\
  \bibinfo {author} {\bibfnamefont {L.}~\bibnamefont {Balents}},\ }\href
  {\doibase 10.1038/nphys2465} {\bibfield  {journal} {\bibinfo  {journal}
  {Nature Physics}\ }\textbf {\bibinfo {volume} {8}},\ \bibinfo {pages} {902}
  (\bibinfo {year} {2012})}\BibitemShut {NoStop}%
\bibitem [{\citenamefont {Levin}\ and\ \citenamefont {Wen}(2006)}]{Wen2006}%
  \BibitemOpen
  \bibfield  {author} {\bibinfo {author} {\bibfnamefont {M.}~\bibnamefont
  {Levin}}\ and\ \bibinfo {author} {\bibfnamefont {X.-G.}\ \bibnamefont
  {Wen}},\ }\href {\doibase 10.1103/PhysRevLett.96.110405} {\bibfield
  {journal} {\bibinfo  {journal} {Phys. Rev. Lett.}\ }\textbf {\bibinfo
  {volume} {96}},\ \bibinfo {pages} {110405} (\bibinfo {year}
  {2006})}\BibitemShut {NoStop}%
\bibitem [{\citenamefont {Hamma}\ \emph
  {et~al.}(2005{\natexlab{a}})\citenamefont {Hamma}, \citenamefont
  {Ionicioiu},\ and\ \citenamefont {Zanardi}}]{HAMMA200522}%
  \BibitemOpen
  \bibfield  {author} {\bibinfo {author} {\bibfnamefont {A.}~\bibnamefont
  {Hamma}}, \bibinfo {author} {\bibfnamefont {R.}~\bibnamefont {Ionicioiu}}, \
  and\ \bibinfo {author} {\bibfnamefont {P.}~\bibnamefont {Zanardi}},\ }\href
  {\doibase https://doi.org/10.1016/j.physleta.2005.01.060} {\bibfield
  {journal} {\bibinfo  {journal} {Physics Letters A}\ }\textbf {\bibinfo
  {volume} {337}},\ \bibinfo {pages} {22} (\bibinfo {year}
  {2005}{\natexlab{a}})}\BibitemShut {NoStop}%
\bibitem [{\citenamefont {Hamma}\ \emph
  {et~al.}(2005{\natexlab{b}})\citenamefont {Hamma}, \citenamefont
  {Ionicioiu},\ and\ \citenamefont {Zanardi}}]{Hamma2005}%
  \BibitemOpen
  \bibfield  {author} {\bibinfo {author} {\bibfnamefont {A.}~\bibnamefont
  {Hamma}}, \bibinfo {author} {\bibfnamefont {R.}~\bibnamefont {Ionicioiu}}, \
  and\ \bibinfo {author} {\bibfnamefont {P.}~\bibnamefont {Zanardi}},\ }\href
  {\doibase 10.1103/PhysRevA.71.022315} {\bibfield  {journal} {\bibinfo
  {journal} {Phys. Rev. A}\ }\textbf {\bibinfo {volume} {71}},\ \bibinfo
  {pages} {022315} (\bibinfo {year} {2005}{\natexlab{b}})}\BibitemShut
  {NoStop}%
\bibitem [{\citenamefont {Zeng}\ \emph {et~al.}(2019)\citenamefont {Zeng},
  \citenamefont {Chen}, \citenamefont {Zhou},\ and\ \citenamefont
  {Wen}}]{Zeng2019}%
  \BibitemOpen
  \bibfield  {author} {\bibinfo {author} {\bibfnamefont {B.}~\bibnamefont
  {Zeng}}, \bibinfo {author} {\bibfnamefont {X.}~\bibnamefont {Chen}}, \bibinfo
  {author} {\bibfnamefont {D.-L.}\ \bibnamefont {Zhou}}, \ and\ \bibinfo
  {author} {\bibfnamefont {X.-G.}\ \bibnamefont {Wen}},\ }\enquote {\bibinfo
  {title} {Gapped quantum systems and entanglement area law},}\ in\ \href
  {\doibase 10.1007/978-1-4939-9084-9_5} {\emph {\bibinfo {booktitle} {Quantum
  Information Meets Quantum Matter: From Quantum Entanglement to Topological
  Phases of Many-Body Systems}}}\ (\bibinfo  {publisher} {Springer New York},\
  \bibinfo {address} {New York, NY},\ \bibinfo {year} {2019})\ pp.\ \bibinfo
  {pages} {115--153}\BibitemShut {NoStop}%
\bibitem [{\citenamefont {Castelnovo}(2014)}]{Castelnovo2014}%
  \BibitemOpen
  \bibfield  {author} {\bibinfo {author} {\bibfnamefont {C.}~\bibnamefont
  {Castelnovo}},\ }\href {\doibase 10.1103/PhysRevA.89.042333} {\bibfield
  {journal} {\bibinfo  {journal} {Phys. Rev. A}\ }\textbf {\bibinfo {volume}
  {89}},\ \bibinfo {pages} {042333} (\bibinfo {year} {2014})}\BibitemShut
  {NoStop}%
\bibitem [{\citenamefont {Kitaev}(2006)}]{KITAEV20062}%
  \BibitemOpen
  \bibfield  {author} {\bibinfo {author} {\bibfnamefont {A.}~\bibnamefont
  {Kitaev}},\ }\href {\doibase https://doi.org/10.1016/j.aop.2005.10.005}
  {\bibfield  {journal} {\bibinfo  {journal} {Annals of Physics}\ }\textbf
  {\bibinfo {volume} {321}},\ \bibinfo {pages} {2} (\bibinfo {year} {2006})},\
  \bibinfo {note} {january Special Issue}\BibitemShut {NoStop}%
\bibitem [{\citenamefont {Knolle}\ and\ \citenamefont
  {Moessner}(2019)}]{Knolle2019}%
  \BibitemOpen
  \bibfield  {author} {\bibinfo {author} {\bibfnamefont {J.}~\bibnamefont
  {Knolle}}\ and\ \bibinfo {author} {\bibfnamefont {R.}~\bibnamefont
  {Moessner}},\ }\href {\doibase 10.1146/annurev-conmatphys-031218-013401}
  {\bibfield  {journal} {\bibinfo  {journal} {Annual Review of Condensed Matter
  Physics}\ }\textbf {\bibinfo {volume} {10}},\ \bibinfo {pages} {451}
  (\bibinfo {year} {2019})}\BibitemShut {NoStop}%
\bibitem [{\citenamefont {Savary}\ and\ \citenamefont
  {Balents}(2017)}]{Blents1}%
  \BibitemOpen
  \bibfield  {author} {\bibinfo {author} {\bibfnamefont {L.}~\bibnamefont
  {Savary}}\ and\ \bibinfo {author} {\bibfnamefont {L.}~\bibnamefont
  {Balents}},\ }\href {http://stacks.iop.org/0034-4885/80/i=1/a=016502}
  {\bibfield  {journal} {\bibinfo  {journal} {Reports on Progress in Physics}\
  }\textbf {\bibinfo {volume} {80}},\ \bibinfo {pages} {016502} (\bibinfo
  {year} {2017})}\BibitemShut {NoStop}%
\bibitem [{\citenamefont {Zhou}\ \emph {et~al.}(2017)\citenamefont {Zhou},
  \citenamefont {Kanoda},\ and\ \citenamefont {Ng}}]{ZhouRMP}%
  \BibitemOpen
  \bibfield  {author} {\bibinfo {author} {\bibfnamefont {Y.}~\bibnamefont
  {Zhou}}, \bibinfo {author} {\bibfnamefont {K.}~\bibnamefont {Kanoda}}, \ and\
  \bibinfo {author} {\bibfnamefont {T.-K.}\ \bibnamefont {Ng}},\ }\href
  {\doibase 10.1103/RevModPhys.89.025003} {\bibfield  {journal} {\bibinfo
  {journal} {Rev. Mod. Phys.}\ }\textbf {\bibinfo {volume} {89}},\ \bibinfo
  {pages} {025003} (\bibinfo {year} {2017})},\ \bibinfo {note} {see also
  references therein.}\BibitemShut {Stop}%
\bibitem [{\citenamefont {Kasahara}\ \emph {et~al.}(2018)\citenamefont
  {Kasahara}, \citenamefont {Ohnishi}, \citenamefont {Mizukami}, \citenamefont
  {Tanaka}, \citenamefont {Ma}, \citenamefont {Sugii}, \citenamefont {Kurita},
  \citenamefont {Tanaka}, \citenamefont {Nasu}, \citenamefont {Motome},
  \citenamefont {Shibauchi},\ and\ \citenamefont {Matsuda}}]{Kasahara2018}%
  \BibitemOpen
  \bibfield  {author} {\bibinfo {author} {\bibfnamefont {Y.}~\bibnamefont
  {Kasahara}}, \bibinfo {author} {\bibfnamefont {T.}~\bibnamefont {Ohnishi}},
  \bibinfo {author} {\bibfnamefont {Y.}~\bibnamefont {Mizukami}}, \bibinfo
  {author} {\bibfnamefont {O.}~\bibnamefont {Tanaka}}, \bibinfo {author}
  {\bibfnamefont {S.}~\bibnamefont {Ma}}, \bibinfo {author} {\bibfnamefont
  {K.}~\bibnamefont {Sugii}}, \bibinfo {author} {\bibfnamefont
  {N.}~\bibnamefont {Kurita}}, \bibinfo {author} {\bibfnamefont
  {H.}~\bibnamefont {Tanaka}}, \bibinfo {author} {\bibfnamefont
  {J.}~\bibnamefont {Nasu}}, \bibinfo {author} {\bibfnamefont {Y.}~\bibnamefont
  {Motome}}, \bibinfo {author} {\bibfnamefont {T.}~\bibnamefont {Shibauchi}}, \
  and\ \bibinfo {author} {\bibfnamefont {Y.}~\bibnamefont {Matsuda}},\ }\href
  {\doibase 10.1038/s41586-018-0274-0} {\bibfield  {journal} {\bibinfo
  {journal} {Nature}\ }\textbf {\bibinfo {volume} {559}},\ \bibinfo {pages}
  {227} (\bibinfo {year} {2018})}\BibitemShut {NoStop}%
\bibitem [{\citenamefont {Arakawa}\ and\ \citenamefont
  {Yonemitsu}(2021)}]{Arakawa2021}%
  \BibitemOpen
  \bibfield  {author} {\bibinfo {author} {\bibfnamefont {N.}~\bibnamefont
  {Arakawa}}\ and\ \bibinfo {author} {\bibfnamefont {K.}~\bibnamefont
  {Yonemitsu}},\ }\href {\doibase 10.1103/PhysRevB.103.L100408} {\bibfield
  {journal} {\bibinfo  {journal} {Phys. Rev. B}\ }\textbf {\bibinfo {volume}
  {103}},\ \bibinfo {pages} {L100408} (\bibinfo {year} {2021})}\BibitemShut
  {NoStop}%
\bibitem [{\citenamefont {Levin}\ and\ \citenamefont {Wen}(2003)}]{Wen2003}%
  \BibitemOpen
  \bibfield  {author} {\bibinfo {author} {\bibfnamefont {M.}~\bibnamefont
  {Levin}}\ and\ \bibinfo {author} {\bibfnamefont {X.-G.}\ \bibnamefont
  {Wen}},\ }\href {\doibase 10.1103/PhysRevB.67.245316} {\bibfield  {journal}
  {\bibinfo  {journal} {Phys. Rev. B}\ }\textbf {\bibinfo {volume} {67}},\
  \bibinfo {pages} {245316} (\bibinfo {year} {2003})}\BibitemShut {NoStop}%
\bibitem [{\citenamefont {Nanda}\ \emph {et~al.}(2021)\citenamefont {Nanda},
  \citenamefont {Agarwala},\ and\ \citenamefont {Bhattacharjee}}]{Adhip2021}%
  \BibitemOpen
  \bibfield  {author} {\bibinfo {author} {\bibfnamefont {A.}~\bibnamefont
  {Nanda}}, \bibinfo {author} {\bibfnamefont {A.}~\bibnamefont {Agarwala}}, \
  and\ \bibinfo {author} {\bibfnamefont {S.}~\bibnamefont {Bhattacharjee}},\
  }\href {\doibase 10.1103/PhysRevB.104.195115} {\bibfield  {journal} {\bibinfo
   {journal} {Phys. Rev. B}\ }\textbf {\bibinfo {volume} {104}},\ \bibinfo
  {pages} {195115} (\bibinfo {year} {2021})}\BibitemShut {NoStop}%
\bibitem [{\citenamefont {Baskaran}\ \emph {et~al.}(2007)\citenamefont
  {Baskaran}, \citenamefont {Mandal},\ and\ \citenamefont
  {Shankar}}]{Baskaran2007}%
  \BibitemOpen
  \bibfield  {author} {\bibinfo {author} {\bibfnamefont {G.}~\bibnamefont
  {Baskaran}}, \bibinfo {author} {\bibfnamefont {S.}~\bibnamefont {Mandal}}, \
  and\ \bibinfo {author} {\bibfnamefont {R.}~\bibnamefont {Shankar}},\ }\href
  {\doibase 10.1103/PhysRevLett.98.247201} {\bibfield  {journal} {\bibinfo
  {journal} {Phys. Rev. Lett.}\ }\textbf {\bibinfo {volume} {98}},\ \bibinfo
  {pages} {247201} (\bibinfo {year} {2007})}\BibitemShut {NoStop}%
\bibitem [{\citenamefont {Wang}\ \emph {et~al.}(2010)\citenamefont {Wang},
  \citenamefont {Ma}, \citenamefont {Gu},\ and\ \citenamefont
  {Lin}}]{Wang2010}%
  \BibitemOpen
  \bibfield  {author} {\bibinfo {author} {\bibfnamefont {Z.}~\bibnamefont
  {Wang}}, \bibinfo {author} {\bibfnamefont {T.}~\bibnamefont {Ma}}, \bibinfo
  {author} {\bibfnamefont {S.-J.}\ \bibnamefont {Gu}}, \ and\ \bibinfo {author}
  {\bibfnamefont {H.-Q.}\ \bibnamefont {Lin}},\ }\href {\doibase
  10.1103/PhysRevA.81.062350} {\bibfield  {journal} {\bibinfo  {journal} {Phys.
  Rev. A}\ }\textbf {\bibinfo {volume} {81}},\ \bibinfo {pages} {062350}
  (\bibinfo {year} {2010})}\BibitemShut {NoStop}%
\bibitem [{\citenamefont {Lahtinen}\ \emph {et~al.}(2012)\citenamefont
  {Lahtinen}, \citenamefont {Ludwig}, \citenamefont {Pachos},\ and\
  \citenamefont {Trebst}}]{Lahtinen2012}%
  \BibitemOpen
  \bibfield  {author} {\bibinfo {author} {\bibfnamefont {V.}~\bibnamefont
  {Lahtinen}}, \bibinfo {author} {\bibfnamefont {A.~W.~W.}\ \bibnamefont
  {Ludwig}}, \bibinfo {author} {\bibfnamefont {J.~K.}\ \bibnamefont {Pachos}},
  \ and\ \bibinfo {author} {\bibfnamefont {S.}~\bibnamefont {Trebst}},\ }\href
  {\doibase 10.1103/PhysRevB.86.075115} {\bibfield  {journal} {\bibinfo
  {journal} {Phys. Rev. B}\ }\textbf {\bibinfo {volume} {86}},\ \bibinfo
  {pages} {075115} (\bibinfo {year} {2012})}\BibitemShut {NoStop}%
\bibitem [{\citenamefont {von L\"upke}\ \emph {et~al.}(2020)\citenamefont {von
  L\"upke}, \citenamefont {Beaudoin}, \citenamefont {Norris}, \citenamefont
  {Sung}, \citenamefont {Winik}, \citenamefont {Qiu}, \citenamefont
  {Kjaergaard}, \citenamefont {Kim}, \citenamefont {Yoder}, \citenamefont
  {Gustavsson}, \citenamefont {Viola},\ and\ \citenamefont
  {Oliver}}]{Lupke2020}%
  \BibitemOpen
  \bibfield  {author} {\bibinfo {author} {\bibfnamefont {U.}~\bibnamefont {von
  L\"upke}}, \bibinfo {author} {\bibfnamefont {F.}~\bibnamefont {Beaudoin}},
  \bibinfo {author} {\bibfnamefont {L.~M.}\ \bibnamefont {Norris}}, \bibinfo
  {author} {\bibfnamefont {Y.}~\bibnamefont {Sung}}, \bibinfo {author}
  {\bibfnamefont {R.}~\bibnamefont {Winik}}, \bibinfo {author} {\bibfnamefont
  {J.~Y.}\ \bibnamefont {Qiu}}, \bibinfo {author} {\bibfnamefont
  {M.}~\bibnamefont {Kjaergaard}}, \bibinfo {author} {\bibfnamefont
  {D.}~\bibnamefont {Kim}}, \bibinfo {author} {\bibfnamefont {J.}~\bibnamefont
  {Yoder}}, \bibinfo {author} {\bibfnamefont {S.}~\bibnamefont {Gustavsson}},
  \bibinfo {author} {\bibfnamefont {L.}~\bibnamefont {Viola}}, \ and\ \bibinfo
  {author} {\bibfnamefont {W.~D.}\ \bibnamefont {Oliver}},\ }\href {\doibase
  10.1103/PRXQuantum.1.010305} {\bibfield  {journal} {\bibinfo  {journal} {PRX
  Quantum}\ }\textbf {\bibinfo {volume} {1}},\ \bibinfo {pages} {010305}
  (\bibinfo {year} {2020})}\BibitemShut {NoStop}%
\bibitem [{\citenamefont {Brydges}\ \emph {et~al.}(2019)\citenamefont
  {Brydges}, \citenamefont {Elben}, \citenamefont {Jurcevic}, \citenamefont
  {Vermersch}, \citenamefont {Maier}, \citenamefont {Lanyon}, \citenamefont
  {Zoller}, \citenamefont {Blatt},\ and\ \citenamefont {Roos}}]{Brydges2019}%
  \BibitemOpen
  \bibfield  {author} {\bibinfo {author} {\bibfnamefont {T.}~\bibnamefont
  {Brydges}}, \bibinfo {author} {\bibfnamefont {A.}~\bibnamefont {Elben}},
  \bibinfo {author} {\bibfnamefont {P.}~\bibnamefont {Jurcevic}}, \bibinfo
  {author} {\bibfnamefont {B.}~\bibnamefont {Vermersch}}, \bibinfo {author}
  {\bibfnamefont {C.}~\bibnamefont {Maier}}, \bibinfo {author} {\bibfnamefont
  {B.~P.}\ \bibnamefont {Lanyon}}, \bibinfo {author} {\bibfnamefont
  {P.}~\bibnamefont {Zoller}}, \bibinfo {author} {\bibfnamefont
  {R.}~\bibnamefont {Blatt}}, \ and\ \bibinfo {author} {\bibfnamefont {C.~F.}\
  \bibnamefont {Roos}},\ }\href {\doibase 10.1126/science.aau4963} {\bibfield
  {journal} {\bibinfo  {journal} {Science}\ }\textbf {\bibinfo {volume}
  {364}},\ \bibinfo {pages} {260} (\bibinfo {year} {2019})},\ \Eprint
  {http://arxiv.org/abs/https://www.science.org/doi/pdf/10.1126/science.aau4963}
  {https://www.science.org/doi/pdf/10.1126/science.aau4963} \BibitemShut
  {NoStop}%
\bibitem [{\citenamefont {Vermersch}\ \emph {et~al.}(2019)\citenamefont
  {Vermersch}, \citenamefont {Elben}, \citenamefont {Sieberer}, \citenamefont
  {Yao},\ and\ \citenamefont {Zoller}}]{Vermersch2019}%
  \BibitemOpen
  \bibfield  {author} {\bibinfo {author} {\bibfnamefont {B.}~\bibnamefont
  {Vermersch}}, \bibinfo {author} {\bibfnamefont {A.}~\bibnamefont {Elben}},
  \bibinfo {author} {\bibfnamefont {L.~M.}\ \bibnamefont {Sieberer}}, \bibinfo
  {author} {\bibfnamefont {N.~Y.}\ \bibnamefont {Yao}}, \ and\ \bibinfo
  {author} {\bibfnamefont {P.}~\bibnamefont {Zoller}},\ }\href {\doibase
  10.1103/PhysRevX.9.021061} {\bibfield  {journal} {\bibinfo  {journal} {Phys.
  Rev. X}\ }\textbf {\bibinfo {volume} {9}},\ \bibinfo {pages} {021061}
  (\bibinfo {year} {2019})}\BibitemShut {NoStop}%
\bibitem [{\citenamefont {Satzinger}\ \emph {et~al.}(2021)\citenamefont
  {Satzinger}, \citenamefont {Liu}, \citenamefont {Smith}, \citenamefont
  {Knapp}, \citenamefont {Newman}, \citenamefont {Jones}, \citenamefont {Chen},
  \citenamefont {Quintana}, \citenamefont {Mi}, \citenamefont {Dunsworth},
  \citenamefont {Gidney}, \citenamefont {Aleiner}, \citenamefont {Arute},
  \citenamefont {Arya}, \citenamefont {Atalaya}, \citenamefont {Babbush},
  \citenamefont {Bardin}, \citenamefont {Barends}, \citenamefont {Basso},
  \citenamefont {Bengtsson}, \citenamefont {Bilmes}, \citenamefont {Broughton},
  \citenamefont {Buckley}, \citenamefont {Buell}, \citenamefont {Burkett},
  \citenamefont {Bushnell}, \citenamefont {Chiaro}, \citenamefont {Collins},
  \citenamefont {Courtney}, \citenamefont {Demura}, \citenamefont {Derk},
  \citenamefont {Eppens}, \citenamefont {Erickson}, \citenamefont {Faoro},
  \citenamefont {Farhi}, \citenamefont {Fowler}, \citenamefont {Foxen},
  \citenamefont {Giustina}, \citenamefont {Greene}, \citenamefont {Gross},
  \citenamefont {Harrigan}, \citenamefont {Harrington}, \citenamefont {Hilton},
  \citenamefont {Hong}, \citenamefont {Huang}, \citenamefont {Huggins},
  \citenamefont {Ioffe}, \citenamefont {Isakov}, \citenamefont {Jeffrey},
  \citenamefont {Jiang}, \citenamefont {Kafri}, \citenamefont {Kechedzhi},
  \citenamefont {Khattar}, \citenamefont {Kim}, \citenamefont {Klimov},
  \citenamefont {Korotkov}, \citenamefont {Kostritsa}, \citenamefont
  {Landhuis}, \citenamefont {Laptev}, \citenamefont {Locharla}, \citenamefont
  {Lucero}, \citenamefont {Martin}, \citenamefont {McClean}, \citenamefont
  {McEwen}, \citenamefont {Miao}, \citenamefont {Mohseni}, \citenamefont
  {Montazeri}, \citenamefont {Mruczkiewicz}, \citenamefont {Mutus},
  \citenamefont {Naaman}, \citenamefont {Neeley}, \citenamefont {Neill},
  \citenamefont {Niu}, \citenamefont {O’Brien}, \citenamefont {Opremcak},
  \citenamefont {Pató}, \citenamefont {Petukhov}, \citenamefont {Rubin},
  \citenamefont {Sank}, \citenamefont {Shvarts}, \citenamefont {Strain},
  \citenamefont {Szalay}, \citenamefont {Villalonga}, \citenamefont {White},
  \citenamefont {Yao}, \citenamefont {Yeh}, \citenamefont {Yoo}, \citenamefont
  {Zalcman}, \citenamefont {Neven}, \citenamefont {Boixo}, \citenamefont
  {Megrant}, \citenamefont {Chen}, \citenamefont {Kelly}, \citenamefont
  {Smelyanskiy}, \citenamefont {Kitaev}, \citenamefont {Knap}, \citenamefont
  {Pollmann},\ and\ \citenamefont {Roushan}}]{Satzinger2021}%
  \BibitemOpen
  \bibfield  {author} {\bibinfo {author} {\bibfnamefont {K.~J.}\ \bibnamefont
  {Satzinger}}, \bibinfo {author} {\bibfnamefont {Y.-J.}\ \bibnamefont {Liu}},
  \bibinfo {author} {\bibfnamefont {A.}~\bibnamefont {Smith}}, \bibinfo
  {author} {\bibfnamefont {C.}~\bibnamefont {Knapp}}, \bibinfo {author}
  {\bibfnamefont {M.}~\bibnamefont {Newman}}, \bibinfo {author} {\bibfnamefont
  {C.}~\bibnamefont {Jones}}, \bibinfo {author} {\bibfnamefont
  {Z.}~\bibnamefont {Chen}}, \bibinfo {author} {\bibfnamefont {C.}~\bibnamefont
  {Quintana}}, \bibinfo {author} {\bibfnamefont {X.}~\bibnamefont {Mi}},
  \bibinfo {author} {\bibfnamefont {A.}~\bibnamefont {Dunsworth}}, \bibinfo
  {author} {\bibfnamefont {C.}~\bibnamefont {Gidney}}, \bibinfo {author}
  {\bibfnamefont {I.}~\bibnamefont {Aleiner}}, \bibinfo {author} {\bibfnamefont
  {F.}~\bibnamefont {Arute}}, \bibinfo {author} {\bibfnamefont
  {K.}~\bibnamefont {Arya}}, \bibinfo {author} {\bibfnamefont {J.}~\bibnamefont
  {Atalaya}}, \bibinfo {author} {\bibfnamefont {R.}~\bibnamefont {Babbush}},
  \bibinfo {author} {\bibfnamefont {J.~C.}\ \bibnamefont {Bardin}}, \bibinfo
  {author} {\bibfnamefont {R.}~\bibnamefont {Barends}}, \bibinfo {author}
  {\bibfnamefont {J.}~\bibnamefont {Basso}}, \bibinfo {author} {\bibfnamefont
  {A.}~\bibnamefont {Bengtsson}}, \bibinfo {author} {\bibfnamefont
  {A.}~\bibnamefont {Bilmes}}, \bibinfo {author} {\bibfnamefont
  {M.}~\bibnamefont {Broughton}}, \bibinfo {author} {\bibfnamefont {B.~B.}\
  \bibnamefont {Buckley}}, \bibinfo {author} {\bibfnamefont {D.~A.}\
  \bibnamefont {Buell}}, \bibinfo {author} {\bibfnamefont {B.}~\bibnamefont
  {Burkett}}, \bibinfo {author} {\bibfnamefont {N.}~\bibnamefont {Bushnell}},
  \bibinfo {author} {\bibfnamefont {B.}~\bibnamefont {Chiaro}}, \bibinfo
  {author} {\bibfnamefont {R.}~\bibnamefont {Collins}}, \bibinfo {author}
  {\bibfnamefont {W.}~\bibnamefont {Courtney}}, \bibinfo {author}
  {\bibfnamefont {S.}~\bibnamefont {Demura}}, \bibinfo {author} {\bibfnamefont
  {A.~R.}\ \bibnamefont {Derk}}, \bibinfo {author} {\bibfnamefont
  {D.}~\bibnamefont {Eppens}}, \bibinfo {author} {\bibfnamefont
  {C.}~\bibnamefont {Erickson}}, \bibinfo {author} {\bibfnamefont
  {L.}~\bibnamefont {Faoro}}, \bibinfo {author} {\bibfnamefont
  {E.}~\bibnamefont {Farhi}}, \bibinfo {author} {\bibfnamefont {A.~G.}\
  \bibnamefont {Fowler}}, \bibinfo {author} {\bibfnamefont {B.}~\bibnamefont
  {Foxen}}, \bibinfo {author} {\bibfnamefont {M.}~\bibnamefont {Giustina}},
  \bibinfo {author} {\bibfnamefont {A.}~\bibnamefont {Greene}}, \bibinfo
  {author} {\bibfnamefont {J.~A.}\ \bibnamefont {Gross}}, \bibinfo {author}
  {\bibfnamefont {M.~P.}\ \bibnamefont {Harrigan}}, \bibinfo {author}
  {\bibfnamefont {S.~D.}\ \bibnamefont {Harrington}}, \bibinfo {author}
  {\bibfnamefont {J.}~\bibnamefont {Hilton}}, \bibinfo {author} {\bibfnamefont
  {S.}~\bibnamefont {Hong}}, \bibinfo {author} {\bibfnamefont {T.}~\bibnamefont
  {Huang}}, \bibinfo {author} {\bibfnamefont {W.~J.}\ \bibnamefont {Huggins}},
  \bibinfo {author} {\bibfnamefont {L.~B.}\ \bibnamefont {Ioffe}}, \bibinfo
  {author} {\bibfnamefont {S.~V.}\ \bibnamefont {Isakov}}, \bibinfo {author}
  {\bibfnamefont {E.}~\bibnamefont {Jeffrey}}, \bibinfo {author} {\bibfnamefont
  {Z.}~\bibnamefont {Jiang}}, \bibinfo {author} {\bibfnamefont
  {D.}~\bibnamefont {Kafri}}, \bibinfo {author} {\bibfnamefont
  {K.}~\bibnamefont {Kechedzhi}}, \bibinfo {author} {\bibfnamefont
  {T.}~\bibnamefont {Khattar}}, \bibinfo {author} {\bibfnamefont
  {S.}~\bibnamefont {Kim}}, \bibinfo {author} {\bibfnamefont {P.~V.}\
  \bibnamefont {Klimov}}, \bibinfo {author} {\bibfnamefont {A.~N.}\
  \bibnamefont {Korotkov}}, \bibinfo {author} {\bibfnamefont {F.}~\bibnamefont
  {Kostritsa}}, \bibinfo {author} {\bibfnamefont {D.}~\bibnamefont {Landhuis}},
  \bibinfo {author} {\bibfnamefont {P.}~\bibnamefont {Laptev}}, \bibinfo
  {author} {\bibfnamefont {A.}~\bibnamefont {Locharla}}, \bibinfo {author}
  {\bibfnamefont {E.}~\bibnamefont {Lucero}}, \bibinfo {author} {\bibfnamefont
  {O.}~\bibnamefont {Martin}}, \bibinfo {author} {\bibfnamefont {J.~R.}\
  \bibnamefont {McClean}}, \bibinfo {author} {\bibfnamefont {M.}~\bibnamefont
  {McEwen}}, \bibinfo {author} {\bibfnamefont {K.~C.}\ \bibnamefont {Miao}},
  \bibinfo {author} {\bibfnamefont {M.}~\bibnamefont {Mohseni}}, \bibinfo
  {author} {\bibfnamefont {S.}~\bibnamefont {Montazeri}}, \bibinfo {author}
  {\bibfnamefont {W.}~\bibnamefont {Mruczkiewicz}}, \bibinfo {author}
  {\bibfnamefont {J.}~\bibnamefont {Mutus}}, \bibinfo {author} {\bibfnamefont
  {O.}~\bibnamefont {Naaman}}, \bibinfo {author} {\bibfnamefont
  {M.}~\bibnamefont {Neeley}}, \bibinfo {author} {\bibfnamefont
  {C.}~\bibnamefont {Neill}}, \bibinfo {author} {\bibfnamefont {M.~Y.}\
  \bibnamefont {Niu}}, \bibinfo {author} {\bibfnamefont {T.~E.}\ \bibnamefont
  {O’Brien}}, \bibinfo {author} {\bibfnamefont {A.}~\bibnamefont {Opremcak}},
  \bibinfo {author} {\bibfnamefont {B.}~\bibnamefont {Pató}}, \bibinfo
  {author} {\bibfnamefont {A.}~\bibnamefont {Petukhov}}, \bibinfo {author}
  {\bibfnamefont {N.~C.}\ \bibnamefont {Rubin}}, \bibinfo {author}
  {\bibfnamefont {D.}~\bibnamefont {Sank}}, \bibinfo {author} {\bibfnamefont
  {V.}~\bibnamefont {Shvarts}}, \bibinfo {author} {\bibfnamefont
  {D.}~\bibnamefont {Strain}}, \bibinfo {author} {\bibfnamefont
  {M.}~\bibnamefont {Szalay}}, \bibinfo {author} {\bibfnamefont
  {B.}~\bibnamefont {Villalonga}}, \bibinfo {author} {\bibfnamefont {T.~C.}\
  \bibnamefont {White}}, \bibinfo {author} {\bibfnamefont {Z.}~\bibnamefont
  {Yao}}, \bibinfo {author} {\bibfnamefont {P.}~\bibnamefont {Yeh}}, \bibinfo
  {author} {\bibfnamefont {J.}~\bibnamefont {Yoo}}, \bibinfo {author}
  {\bibfnamefont {A.}~\bibnamefont {Zalcman}}, \bibinfo {author} {\bibfnamefont
  {H.}~\bibnamefont {Neven}}, \bibinfo {author} {\bibfnamefont
  {S.}~\bibnamefont {Boixo}}, \bibinfo {author} {\bibfnamefont
  {A.}~\bibnamefont {Megrant}}, \bibinfo {author} {\bibfnamefont
  {Y.}~\bibnamefont {Chen}}, \bibinfo {author} {\bibfnamefont {J.}~\bibnamefont
  {Kelly}}, \bibinfo {author} {\bibfnamefont {V.}~\bibnamefont {Smelyanskiy}},
  \bibinfo {author} {\bibfnamefont {A.}~\bibnamefont {Kitaev}}, \bibinfo
  {author} {\bibfnamefont {M.}~\bibnamefont {Knap}}, \bibinfo {author}
  {\bibfnamefont {F.}~\bibnamefont {Pollmann}}, \ and\ \bibinfo {author}
  {\bibfnamefont {P.}~\bibnamefont {Roushan}},\ }\href {\doibase
  10.1126/science.abi8378} {\bibfield  {journal} {\bibinfo  {journal}
  {Science}\ }\textbf {\bibinfo {volume} {374}},\ \bibinfo {pages} {1237}
  (\bibinfo {year} {2021})},\ \Eprint
  {http://arxiv.org/abs/https://www.science.org/doi/pdf/10.1126/science.abi8378}
  {https://www.science.org/doi/pdf/10.1126/science.abi8378} \BibitemShut
  {NoStop}%
\bibitem [{\citenamefont {Verresen}\ \emph {et~al.}(2021)\citenamefont
  {Verresen}, \citenamefont {Lukin},\ and\ \citenamefont
  {Vishwanath}}]{Verresen2021}%
  \BibitemOpen
  \bibfield  {author} {\bibinfo {author} {\bibfnamefont {R.}~\bibnamefont
  {Verresen}}, \bibinfo {author} {\bibfnamefont {M.~D.}\ \bibnamefont {Lukin}},
  \ and\ \bibinfo {author} {\bibfnamefont {A.}~\bibnamefont {Vishwanath}},\
  }\href {\doibase 10.1103/PhysRevX.11.031005} {\bibfield  {journal} {\bibinfo
  {journal} {Phys. Rev. X}\ }\textbf {\bibinfo {volume} {11}},\ \bibinfo
  {pages} {031005} (\bibinfo {year} {2021})}\BibitemShut {NoStop}%
\bibitem [{\citenamefont {Pollmann}\ \emph {et~al.}(2010)\citenamefont
  {Pollmann}, \citenamefont {Turner}, \citenamefont {Berg},\ and\ \citenamefont
  {Oshikawa}}]{Pollmann2010}%
  \BibitemOpen
  \bibfield  {author} {\bibinfo {author} {\bibfnamefont {F.}~\bibnamefont
  {Pollmann}}, \bibinfo {author} {\bibfnamefont {A.~M.}\ \bibnamefont
  {Turner}}, \bibinfo {author} {\bibfnamefont {E.}~\bibnamefont {Berg}}, \ and\
  \bibinfo {author} {\bibfnamefont {M.}~\bibnamefont {Oshikawa}},\ }\href
  {\doibase 10.1103/PhysRevB.81.064439} {\bibfield  {journal} {\bibinfo
  {journal} {Phys. Rev. B}\ }\textbf {\bibinfo {volume} {81}},\ \bibinfo
  {pages} {064439} (\bibinfo {year} {2010})}\BibitemShut {NoStop}%
\end{thebibliography}%
\end{document}